\documentclass[aps,11pt,twoside,showpacs,preprintnumbers,tightenlines,nofootinbib,showpacs,article]{revtex4}
\usepackage{graphicx}
\usepackage[sort&compress]{natbib}
\begin{document}
\newcommand{\eg}{{\it e.g.}}
\newcommand{\etal}{{\it et. al.}}
\newcommand{\ie}{{\it i.e.}}
\newcommand{\be}{\begin{equation}}
\newcommand{\dd}{\displaystyle}
\newcommand{\ee}{\end{equation}}
\newcommand{\bea}{\begin{eqnarray}}
\newcommand{\eea}{\end{eqnarray}}
\newcommand{\bef}{\begin{figure}}
\newcommand{\eef}{\end{figure}}
\newcommand{\bce}{\begin{center}}
\newcommand{\ece}{\end{center}}
\def\lsim{\mathrel{\rlap{\lower4pt\hbox{\hskip1pt$\sim$}}
    \raise1pt\hbox{$<$}}}         
\def\gsim{\mathrel{\rlap{\lower4pt\hbox{\hskip1pt$\sim$}}
    \raise1pt\hbox{$>$}}}         

\title{Effective degrees of freedom and gluon condensation in the high temperature
deconfined phase}
\author{P.~Castorina}
\affiliation{Dipartimento di Fisica, Universit\`a di Catania, Catania,  Italia
and INFN Sezione di Catania, Catania, Italia}

\author{M.~Mannarelli}
\affiliation{Center for Theoretical Physics, Laboratory for Nuclear
Science and Department of Physics
\\Massachusetts Institute of Technology, Cambridge, MA 02139.}
\date{\today}
\begin{abstract}
The Equation of State and the properties of matter in the high
temperature  deconfined phase are analyzed by a  quasiparticle
approach for $T> 1.2~T_c$. In order to fix the parameters of our
model we employ the lattice QCD data of energy density and pressure.
First  we consider the pure SU(3) gluon plasma   and it turns out
that such a system can be described in terms of a gluon condensate
and of gluonic quasiparticles whose effective number of degrees of
freedom and mass decrease with increasing temperature. Then we
analyze   QCD with finite quark masses. In this case the numerical
lattice data for energy density and pressure can be fitted assuming
that the system consists of a mixture of gluon quasiparticles,
fermion quasiparticles, boson correlated pairs (corresponding to
in-medium mesonic states) and  gluon condensate. We find that the
effective number of boson degrees of freedom and the in-medium
fermion masses decrease with increasing temperature. At $T \simeq
1.5 ~T_c$ only the correlated pairs corresponding to the mesonic
nonet survive and they completely disappear at $T \simeq 2 ~T_c$.
The temperature dependence of the velocity of sound of the various
quasiparticles, the effects of the breaking of conformal invariance
and the thermodynamic consistency are discussed in detail.

\end{abstract}
\preprint{MIT-CTP 3805}
   \maketitle

\vfill
\eject

\section{Introduction}

The state of matter at asymptotic high temperature is predicted
unambiguously by Quantum Chromodynamics (QCD). Indeed, it is by now
well-established that,  at sufficiently high temperature $T$, the
hadronic constituents form a deconfined and chirally symmetric
Quark-Gluon Plasma  (QGP). In this regime, due to the asymptotic
freedom of QCD, the QGP is expected to be a weakly interacting gas
of quark- and gluon-quasiparticles.

However, the various heavy-ion experiments at Brookhaven, GSI and
CERN laboratories, devoted to the creation and detection of  new
forms of highly excited matter indicate that,  at the reachable
energy scales, the produced phase  exhibits a strong collective
behavior which is incompatible with a weakly interacting QGP.

In particular, standard perturbative QCD cross sections for quarks
and gluons do not allow for a rapid thermalization and fail to
reproduce, by  hydrodynamical models, the observed  phenomena (e.g.
the so-called elliptic flow). Moreover, the results obtained at the
Relativistic Heavy Ion Collider (RHIC) seem to  show that, at large
temperature and small baryon density,  the new state of matter  can
be described as an almost perfect fluid of strongly interacting
particles.

At present the best available tool to study the non-perturbative
properties of QCD in these extreme conditions  is the numerical
simulation on the lattice. Lattice QCD (lQCD)
\cite{Boyd:1995zg,Boyd:1996bx} gives precise indications on the
deviations  of the  pressure, $p$, and of the energy density,
$\epsilon$, of the system from the Stefan-Boltzman (SB) values of an
ideal gas of quarks and gluons even at temperatures as high as $T
\sim 4~T_c$. This suggests that remnants of the confining
interaction play a non-trivial role in the deconfined phase
\cite{Kaczmarek:2004gv}.

In order to motivate our approach to the deconfined phase of QCD let
us first consider how the properties of matter in the confined phase
change with increasing temperature. It is well known that the
 hadronic content of matter and its dynamical properties
 change dramatically with increasing temperature~
\cite{Hagedorn:1965st,Gerber:1988tt,Bebie:1991ij,Leutwyler:1992mv}.
In particular, for $T=100$ MeV, the largest fraction of particles
($\sim 82 \%$) are pions, a smaller fraction, $\sim 12 \%$, consists
of strange mesons  and only about $6\%$ of particles are in excited
states. The mean free path can be estimated to be  $\sim 15 $ fm and
the volume available per particle is roughly $30$ fm$^3$. In these
conditions hadrons do not overlap and the system can be described as
a mixture of the various hadronic states. With increasing
temperature the density increases and the hadronic content changes.
As an example, for $T=150$ MeV, the excited particles are $\sim
40\%$, the remaining $60\%$ of particles are pions and strange
mesons; the mean free path is $\sim 2$fm and the volume per particle
is $4 $ fm$^3$. Therefore in the confined phase, with increasing
temperature, the percentage of particles in excited states increases
and  close to the transition temperature it is reasonable to  expect
that the density of particles is large with a large fraction in
excited states. Since the mean free path can be of the order of the
average distance between particles the system may have  liquid like
properties.

For temperature larger than the critical temperature, it is not
clear which are the correct degrees of freedom. In any case since
the lattice simulations show that the transition from the confined
phase to the deconfined phase is a smooth crossover
\cite{Aoki:2006br,Aoki:2006we}, we expect that for $T \gtrsim T_c$
mesonic states may survive in the deconfined phase. This conclusion
is confirmed by the surprising results of Refs.
\cite{Karsch:2003jg,Asakawa:2002xj,Asakawa:2003nw} where, by
evaluating the static two point correlation function of a
quark-antiquark pair, it has been shown that correlated $\bar q q$
states survive up to temperatures   $ \simeq 2 T_c$. Moreover it
seems that the dissociation temperature for the correlated states
depends on the mass of the quarks, the higher the mass the higher
the melting temperature \cite{Wong:2004zr,Cabrera:2006wh}. This
implies that a detailed analysis is required to understand the
particle content of matter or, more generally,  the effective
degrees of freedom above $T_c$. The presence of mesonic states in
the deconfined phase, observed in lQCD, has also been discussed by
various authors in the random-phase approximation
\cite{Hatsuda:1985eb} and by employing phenomenological models
\cite{Brown:2003km,Castorina:2005tm,Mannarelli:2005pa,Mannarelli:2005pz}
and has been shown \cite{vanHees:2004gq} that may lead to a rapid
thermalization of  heavy quarks in the QGP.

Since   correlated mesonic states may survive in the deconfined
phase they have to be included in a consistent thermodynamical
description of the system above $T_c$. In our previous paper
\cite{Castorina:2005wi}, we have, indeed,  shown that lQCD data of
pressure, $p$,  and energy density, $\epsilon$, in the range  $T =
1.2 - 2$ $ T_c$, are consistent with a description based on $\bar q,
q, g$ quasiparticles, relatively light boson states and gluon
condensate.

In the present paper we shall review in more details those results
and analyze  in a more general way  the thermodynamics of the system
and the role of the gluon condensate. Due to the short mean free
path, a quasi-particle approach is probably not reliable for
temperature  very close to the transition point, therefore we expect
our results to be valid for temperatures larger than  $\sim
1.2~T_c$.

In   Section II we  discuss   the gluon plasma i.e. pure SU(3) gauge
theory.  The lattice results of the so-called interaction measure,
 indicate  deviations of the system from the conformal symmetric
behavior. We find that part of this deviation is due to the gluon
condensate. However,  the gluon condensate is not enough to fit the
lattice values of the trace anomaly at finite temperature, that
measures the breaking of the conformal symmetry. Therefore there
must be other mechanisms that break the scale invariance of the
system (dynamically) generating a new scale. In Section
\ref{quenched}, where we discuss a quasiparticle model for the gluon
plasma, we assume that such  a mechanism can be described in terms
of massive gluon quasiparticle whose effective number of degrees of
freedom is also temperature dependent.  Since at asymptotic high
temperatures QCD is approximately scale invariant we consistently
find that the effective gluon mass decreases with increasing
temperatures. In Section \ref{QGP} we extend our analysis to the
Quark-Gluon Plasma and show that  a quasiparticle description can
reproduce the lattice results of the energy density and of the
pressure of the QGP. This in turn leads to a constraint on the
number of effective boson degrees of freedom and on the mass of the
fermion quasiparticles. Section \ref{conclusion} is devoted to
conclusions and outlooks.

\section{Analysis of the   Gluon Plasma \label{interaction}}

Lattice simulations of  pressure and energy density provide useful
information regarding the properties of QCD in a large range of
temperatures (for a review see Ref.\,\cite{Miller:2006hr}). One
interesting point is to understand whether the system can be
approximately scale invariant. We know that QCD at zero temperature
is not a conformal invariant theory: the coupling constant depends
on a typical  energy scale, $\Lambda_{QCD}$, and the stress energy
tensor $\Theta^{\mu\nu}$ has a trace anomaly. In this Section we
analyze the pure SU(3) gluon plasma, where it is assumed that the
masses of the fermions are infinitely large, i.e. in the so-called
quenched approximation; the analysis of full QCD with finite quark
masses (unquenched approximation) will be developed in the next
Section.

The trace of the average energy momentum tensor is related to the
energy density, $\epsilon$, and the pressure, $p$, of the system by
the relation: \be \Theta^{\mu}_{\mu}\, =\, \epsilon - 3\,p \,.\ee At
finite temperature,  one can write \cite{Leutwyler:1992cd},
\begin{equation}
\Theta^{\mu}_{\mu}(T) = \epsilon(T) - 3p(T)\,=\, <G^2>_0 -
<G^2>_T\,, \label{anomaly}
\end{equation}
where $G^2 = - \frac{\beta}{g} G_{\mu \nu}^a G^{\mu \nu}_a$ and
$<G^2>_0$ and $<G^2>_T$ are the gluon condensate at zero and at
finite temperature respectively. Note that the energy density
$\epsilon(T)$ and pressure $p(T)$ are consistently normalized to
zero at $T=0$.

Equation (\ref{anomaly}) states that the breaking of the conformal
invariance at finite temperature is only due to the gluon
condensate. We can check this relation  by using different lattice
results. Indeed in Ref.~\cite{D'Elia:1997ne,D'Elia:2002ck}
 the gauge invariant two-point correlation
function of the gauge field strengths have been evaluated on the
lattice by using the correlator method. In particular in
Ref.~\cite{D'Elia:2002ck} the gluon condensate has been evaluated at
finite temperature in both pure gauge and full-QCD.  It turns out
that, at moderate temperatures $T \lesssim 2 T_c $, the
chromo-magnetic component $<G^2>^m$  of the gluon condensate
survives above the deconfining temperature and  is (within the
statistical errors)  temperature independent:
\begin{equation}
 \frac{<G^2>^m_T}{{<G^2>^m_0}} \simeq 1
\label{anomaly2}\,,
\end{equation}
whereas the ratio between the chromo-electric contributions
$<G^2>^e$ at zero and at finite temperature,
\begin{equation}
 \frac{<G^2>^e_T}{<G^2>^e_0}
 = c_e^{\rm qu}(T)
\label{anomaly3}\,,
\end{equation}
 rapidly decreases above the deconfining temperature.

Since a $T=0$ the electric and magnetic terms are equal, defining,
\begin{equation}
\Delta_1(T)\,= \frac{1}{2T^4} <G^2>_0 [1- c_e^{\rm qu}(T)]
\label{Delta1}\,,
\end{equation}
and assuming that in a pure gluon plasma  the conformal invariance
is only broken by the gluon condensate, one has
  \be \,\frac{\epsilon -
3p}{T^4}=\Delta_1(T) \label{anomaly4}\,.\ee

Now considering the lattice results of energy density, $\epsilon_L$,
and pressure, $p_L$, of Ref.~\cite{Boyd:1995zg,Boyd:1996bx} we can
obtain  the ``interaction measure" for a pure SU(3) gluon plasma,
\be \Delta(T)= \frac{\epsilon_L(T) -3 p_L(T)}{T^4}\,, \label{Delta}
\ee and we can compare such expression  with $ \Delta_1(T)$.

\begin{figure}[!th]
\includegraphics[width=2.in,angle=-0]{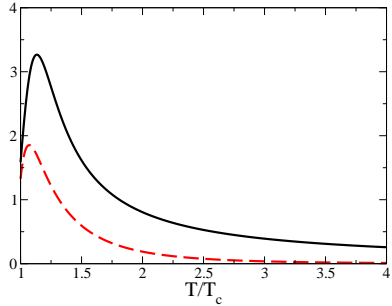}
\caption{(color online) Interaction measure $\Delta(T)$ defined in
Eq.(\ref{Delta}), full line (black online), and contributions of the
gluon condensate  $\Delta_1$, dashed line (red online), defined in
Eq.~(\ref{Delta1}) as a function of $T/T_c$ in the pure SU(3)
Yang-Mills.\vspace{1cm}} \label{intmesureq}
\end{figure}

In Fig.~\ref{intmesureq} we show the plots of $ \Delta(T)$, full
line (black online), and of the $ \Delta_1(T)$, dashed line (red
online). In this plot we are assuming that the
 gluon condensate at zero temperature is given by, $<G^2>_0
 \simeq 0.03$ GeV$^4$ \cite{Shifman:1978bx,Shifman:1978by,Reinders:1984sr}
 and $c_e^{{\rm qu}}(T)$ is obtained by interpolation of the
data obtained in lattice QCD simulations reported in
Ref.~\cite{D'Elia:2002ck}.\vspace*{0.5cm}

It is clear from Fig.~\ref{intmesureq} that the gluon condensate is
not able to describe  the trace anomaly and other contributions must
be present that  stem from interactions and/or masses.

The gluon condensate plays an important dynamical role in QCD.
Indeed, at zero temperature it contributes to a negative
non-perturbative vacuum energy.  In general, a bosonic condensate is
a macroscopically populated state with zero momentum and the
contribution of the modes with $k \neq 0$ is small. For these
reasons,  we will assume  that at any temperature the gluon
condensate does not significantly contribute to the pressure but
does contribute to the energy density of the system.

\begin{figure}[!th]
\includegraphics[width=2.5in,angle=-0]{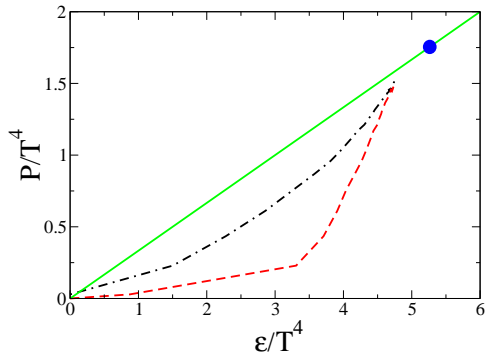}
\caption{(color online) Plots of pressure as a function of energy
density. The diagonal line (green online)  corresponds to the
conformal solution with $p=\epsilon/3$. The dashed line (red online)
 corresponds to the lattice data of energy and pressure, whereas
the dot-dashed (back online) line corresponds to  the quasiparticle
gluons, with pressure equal to $p_L$ and energy $\epsilon_L -
\Delta_1$. The (blue) dot on the diagonal line corresponds to the
ideal SU(3) gas solution.} \label{EP}
\end{figure}

In order to consider  the effect of the gluon condensate on the
energy density we  consider the quantity $\epsilon_L - \Delta_1$. In
Fig.~\ref{EP} we show the plots corresponding to the lattice data of
pressure, $p_L$, versus energy density, $\epsilon_L$, dashed line
(red online) and the data of pressure, $p_L$, versus $\epsilon_L -
\Delta_1$. \vspace*{0.5cm}

The effect of subtracting the gluon condensate is of
 making the relation between energy density and pressure of the gluon system
more similar to the conformal solution, that corresponds to the
diagonal full line $p= \epsilon/3$. However, subtracting the gluon
condensate is not enough for making the system conformal invariant.
Note that for $p=0$ and $\epsilon=0$ the system seems to be
conformal, but this happens because the pressure and energy density
have been normalized to be zero in the confined phase. The (blue)
dot on the diagonal line corresponds to the energy density and
pressure of an ideal gluon gas.

Therefore one can conclude that the trace anomaly at finite
temperature is not saturated by gluon condensation and that the
description of the system requires other dynamical ingredients. The
rigorous analysis in Ref.\cite{Zwanziger:2004np}, where a gluon gas
with physical state space  reduced to the so-called Gribov
fundamental modular region has been considered, gives essentially the
same indication.

In the next sections we shall consider a more phenomenological
approach, based on the introduction of massive gluon quasi-particle,
to describe the difference between trace anomaly and gluon
condensate shown in Figs. \ref{intmesureq} and \ref{EP}.

\subsection{Quasiparticle model for the Gluon Plasma \label{quenched}}

Let us recall some thermodynamical relations. The free energy
density of a system is given by \be f = - \frac{T}{V} {\rm ln}
Z(T,V) \label{freeenergy}\,, \ee where $Z(T,V)$ is the partition
function, and  energy density and pressure are given by,
\begin{eqnarray}
\epsilon&=&\frac{T^2}{V} \frac{\partial {\rm ln} Z(T,V)}{\partial T}\;,
\label{energydens} \\
p&=&T \frac{\partial {\rm ln} Z(T,V)}{\partial V}\;. \label{therm}
\end{eqnarray}
For homogeneous systems, in the thermodynamic limit,  \be \frac{{\rm
ln} Z(T,V)}{V} \simeq \frac{\partial {\rm ln} Z(T,V)}{\partial
V}\,,\ee and therefore the pressure can directly be obtained from
the free energy density, \be p = -f\;. \label{pressure} \ee Using
this relation one can express the entropy density $s$ and the trace
anomaly
 in terms of derivatives of
the pressure with respect to temperature, \be s = {\epsilon + p
\over T}  =  {\partial p \over \partial T } ~,\label{entropy}\ee \be
<\Theta^\mu_\mu>_T={\epsilon - 3p }  =  T^5 {\partial \over
\partial T} (p/T^4) \;. \label{anomaly6}
\ee

Note that for an ideal gas $p \sim T^4$ and the trace of the stress
energy tensor is equal to zero. However, when interactions and/or
masses are present the pressure is not proportional to $T^4$ and one
has to include these effects in  the trace anomaly.

An effective method to take  into account the interaction is by
considering temperature dependent gluon masses and temperature
dependent gluonic degrees of freedom. Therefore in our analysis  we
work in a quasiparticle picture  and all the in-medium effects are
treated, at the mean field level, as an effective mass, $M_g(T)$,
and as an effective number of degrees of freedom, $D_g(T)$, for the
gluons. The values of these parameters will depend on the
temperature and this dependence will be determined by fitting the
lattice data of pressure and energy density where the gluon
condensate contribution will be  properly taken into account.
However, temperature dependent parameters and also the presence of
the gluon condensate require that the thermodynamical consistency
must be carefully checked. This point is discussed in details in the
Appendix A.

According to the previous discussion, let us rewrite
Eq.~(\ref{anomaly}) for a gluon plasma as follows: \be
<\Theta^{\mu}_{\mu}>_T = \epsilon - 3p\,= <G^2>_0-<G^2>_T  +
\Delta\Theta_g \label{anomalyglue}\, , \ee where $\Delta\Theta_g$
represents the contribution to the trace anomaly due to gluon
quasiparticles.

As discussed, the gluon condensate is an  important dynamical
ingredient and we will assume that it contributes to the energy
density but not to the pressure of  the system. Therefore in our
quasiparticle picture, where the contributions to the
thermodynamical quantities come from gluon condensation  and gluons
with in medium mass $M_g(T)$, the pressure and energy density of the
system are given by, \be p = p_g \hspace{2cm} \epsilon = \epsilon_g
+ \epsilon_{\rm con}^{\rm qu}\;, \label{pe}\ee  where, \be
\epsilon_{\rm con}^{\rm qu} = T^4 \Delta_1(T) = \frac{1}{2} <G^2>_0
[1 - c_e^{\rm qu}(T)] \,,\ee is contribution to the energy density
due to the gluon condensate. The quasiparticle contributions to
pressure and energy density are respectively, \bea p_g(T) &=& T
 D_g(T)\int \frac{d^3 k}{(2 \pi)^3} \log(1 - e^{-\omega_g/T})^{-1}\,
 =\, D_g(T) \tilde p_g(T)\,, \nonumber\\
\epsilon_g(T) &=&  D_g(T)\int \!\!\!\frac{d^3 k}{(2 \pi)^3}
\frac{\omega_g}{e^{\omega_g/T} - 1}\,=\,  D_g(T) \tilde
\epsilon_g(T) \label{pressenergyg}\, ,\eea with $D_g(T)$ the
temperature dependent number of gluon degrees of freedom, $\tilde
p_g(T)$ and $\tilde \epsilon_g(T)$ the gluonic pressure and energy
density per degree of freedom, and \be\omega_g(k,T)=\sqrt{k^2 +
M_g(T)^2}\label{gluondisp}\,,\ee is the gluon dispersion law.

\subsection{Determination of $M_g$ and $D_g$ \label{mgdg}}

In order to obtain the functions $D_g(T)$  and $M_g(T)$ we will
employ the   data of the pressure and energy density  obtained in
lattice simulations reported in
Refs.~\cite{Boyd:1995zg,Boyd:1996bx}. We obtain $D_g(T)$ and
$M_g(T)$ substituting  \be p_L(T) = p(T) \hspace{3cm} \epsilon_L(T)
= \epsilon(T)\,, \ee in the  Equations (\ref{pe}) and
(\ref{pressenergyg}), and solving for the   effective mass of the
gluons and the effective number of the degrees of freedom. In
particular the gluonic mass can be obtained solving the equation,
\be \frac{\epsilon_L(T)-\epsilon_{\rm con}^{\rm qu}(T)}{p_L(T)} =
\frac{\tilde \epsilon_g(T)}{\tilde p_g(T)}\;,\label{glue1}\ee and
the effective number of gluonic degrees of freedom can be  obtained
from the relation, \be D_g(T) = \frac{p_L(T)}{\tilde p_g(T)}
\label{glue2}\;,\ee where  the solution for the effective gluon mass
obtained solving Eq.~(\ref{glue1})  has been plugged in $\tilde
p_g(T)$.


Before presenting our results for  $D_g(T)$ and $M_g(T)$, it is
useful to clarify some points regarding the range of values of the
temperature where our results are reliable. Let us consider the
equation \be \,\Delta(T) = \Delta_1(T)+ \frac{D_g(T)(\tilde
\epsilon_g-3 \tilde p_g)}{T^4}\,, \ee which at the critical
temperature  implies \be
 \frac{D_g(T_c)(\tilde \epsilon_ g-3 \tilde p_g)|_{T_c}}{T_c^4}=\Delta(T_c) - \Delta_1(T_c) = d_c
\,.\label{subtraction}\ee

From this equation and from Eq.~(\ref{glue1}) one can see that close
to $T_c$ the values of $M_g(T)$ and $D_g(T)$ will strongly depend on
the values of the lattice data.

Indeed, for $T$ close to $T_c$ there are two solutions of
Eq.~(\ref{glue1}): $M_g(T_c) \rightarrow 0$ (corresponding to the
saturation of the trace anomaly with the gluon condensate) but also
$M_g(T_c) \rightarrow \infty$. Moreover, for $T \rightarrow T_c$, we
have that $d_c \rightarrow 0$ and the actual value for $D_g(T)$ will
 depend on the ratio between two small quantities.  In
particular we find that when the gluon mass diverges also $D_g(T_c)
$ diverges, unless $d_c$ is exactly zero at $T_c$. In any case,
 as previously discussed, a quasi-particle approach is
unreliable at the critical point and for these reasons
 our numerical analysis  will be limited to the range $\sim
1.2 - 4$ $T_c$.

\begin{figure}[!th]
 \hspace{1cm}
\includegraphics[width=2.5in,angle=-0]{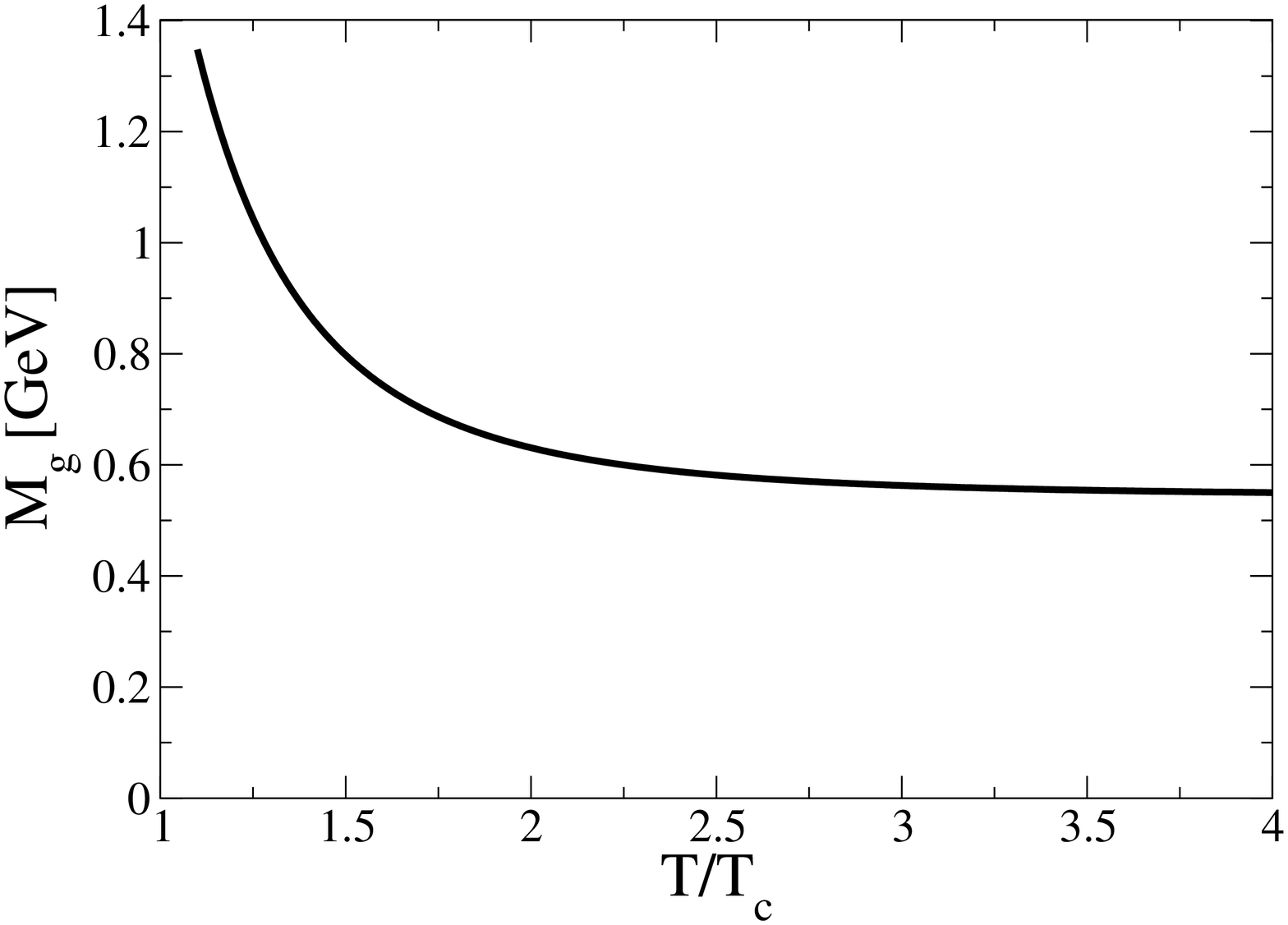}
\includegraphics[width=2.5in,angle=-0]{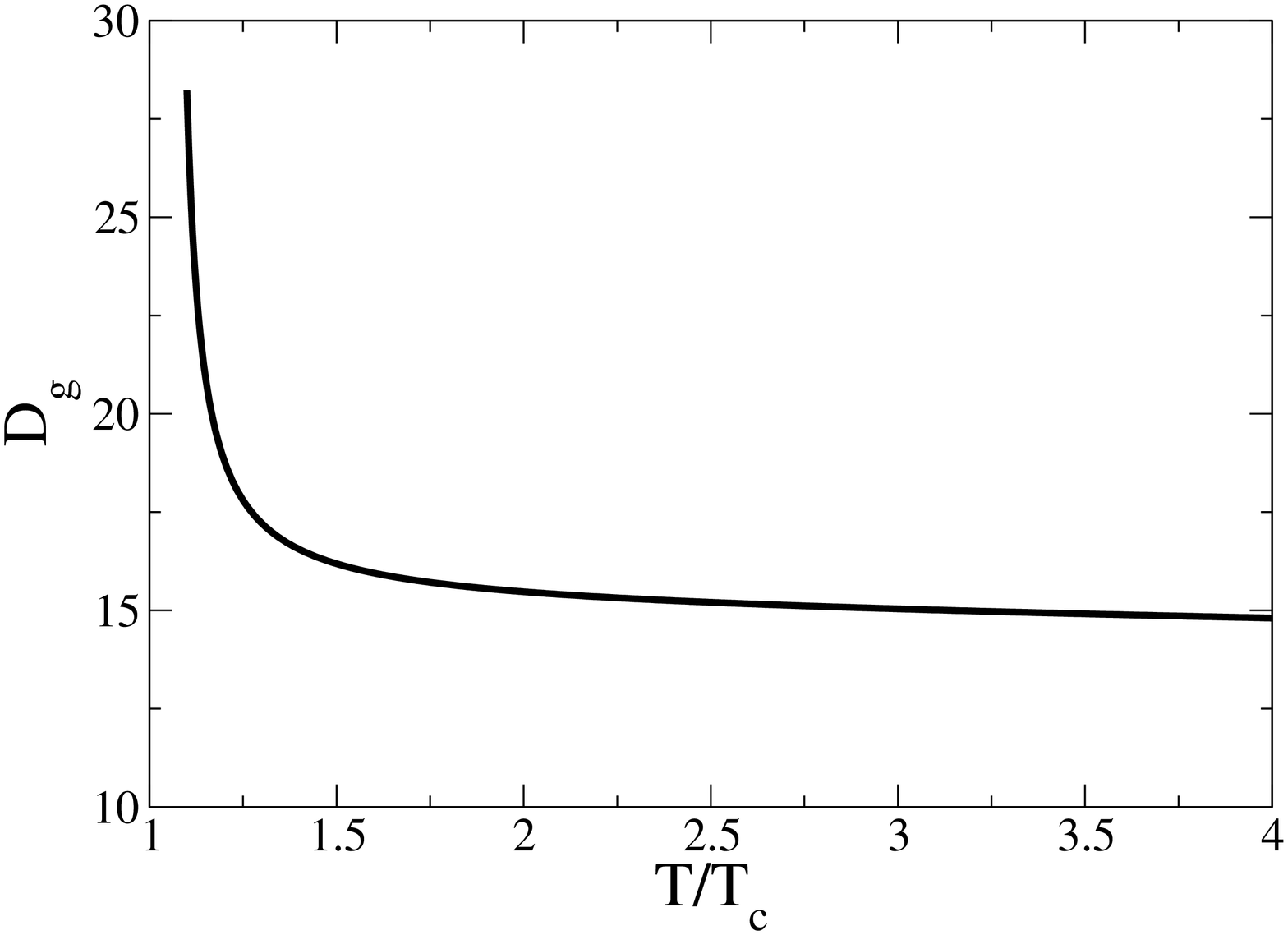}
\caption{Left panel: Effective  mass of the gluonic degrees of
freedom in the quenched approximation as a function of the
temperature. Right panel: Effective number of  degrees of freedom of
the gluonic quasiparticles in the quenched approximation as a
function of the temperature.} \label{figglue}
\end{figure}

From the solutions of Eqs.~(\ref{glue1}) and (\ref{glue2})  we
obtained the values of $M_g(T)$ and $D_g(T)$   reported in
Fig.~\ref{figglue}.\vspace{0.5cm}

Both the effective number of degrees of freedom and the effective
mass monotonically decrease with increasing temperature. Therefore,
in our quasiparticle approach to  the deconfined gluon plasma,
gluons are  propagating quasiparticles with  a  mass and an
effective number of  modes that decrease with increasing
temperature. The reduction of the number of the gluonic modes with
temperature  is not  new  and is physically due to the fact that the
 spectrum of the gluons  depends on the temperature. For momenta $k \lesssim g T$ not only
the transverse components are present, but also the longitudinal
ones. On the other hand, at high momenta $k>>gT$ the longitudinal
modes are not relevant \cite{Pisarski:1989cs} because the
corresponding pole in the propagator becomes exponentially small.
Since the equation of state is dominated by particles with momenta
$k \sim T$ one can expect that at low temperatures the longitudinal
modes are excited, whereas in the high temperature regime the
contribution of the longitudinal poles becomes negligible.  In
principle we should  treat the longitudinal  modes employing a
different dispersion law, however for the sake of simplicity we have
assumed that all modes are degenerate in mass. This is a first
possible ingredient of a more complete dynamical description of the
system  that may lead also to avoid the divergent behavior of the
effective number of degrees at the transition temperature. A second
possible motivation for the sudden increase of the  number of
degrees of freedom close to $T_c$ is that other quasiparticles, like
glueballs, may be present (analogously to the fermionic correlated
states that survive above $T_c$). As a matter of fact it has
recently been pointed out in Ref. \cite{Kochelev:2006sx}  that the
mass of scalar and pseudoscalar glueballs in the deconfined phase is
much less than in the confined phase. Therefore one can expect that
these states significantly contribute to the various thermodynamical
quantities.

A third possibility is that the dispersion relation in
Eq.(\ref{dispersion}) is strongly modified in the infrared region as
in Ref.\cite{Zwanziger:2004np}.

In any case, even with our very simple model, we obtain that  at
high temperatures the number of the gluonic modes is roughly 15, not
far from the expected  result 16, that corresponds to the number of
transverse modes.

\vspace{1cm}

\begin{figure}[!th]
\includegraphics[width=2.5in,angle=-0]{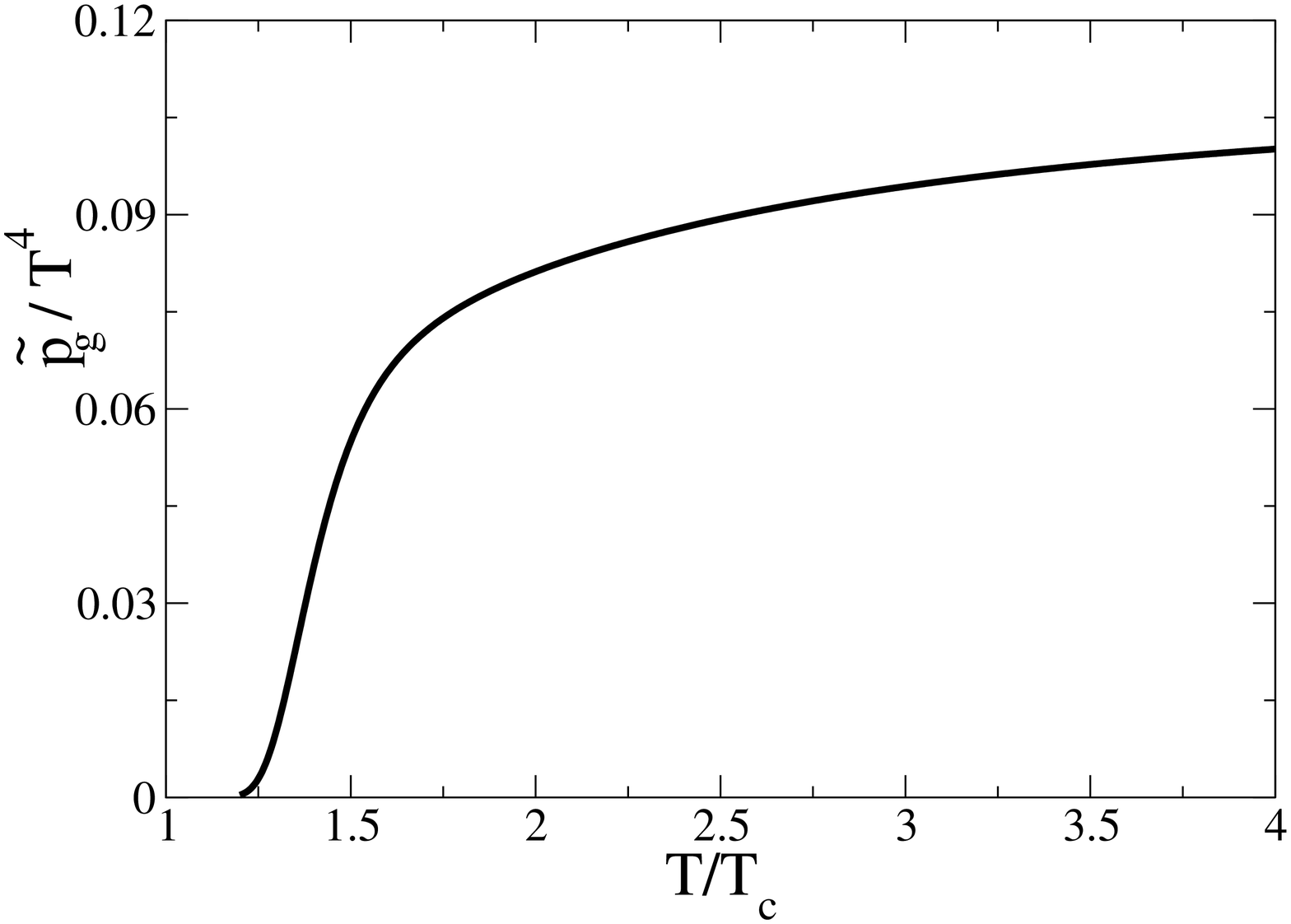} \hspace{1cm}
\includegraphics[width=2.5in,angle=-0]{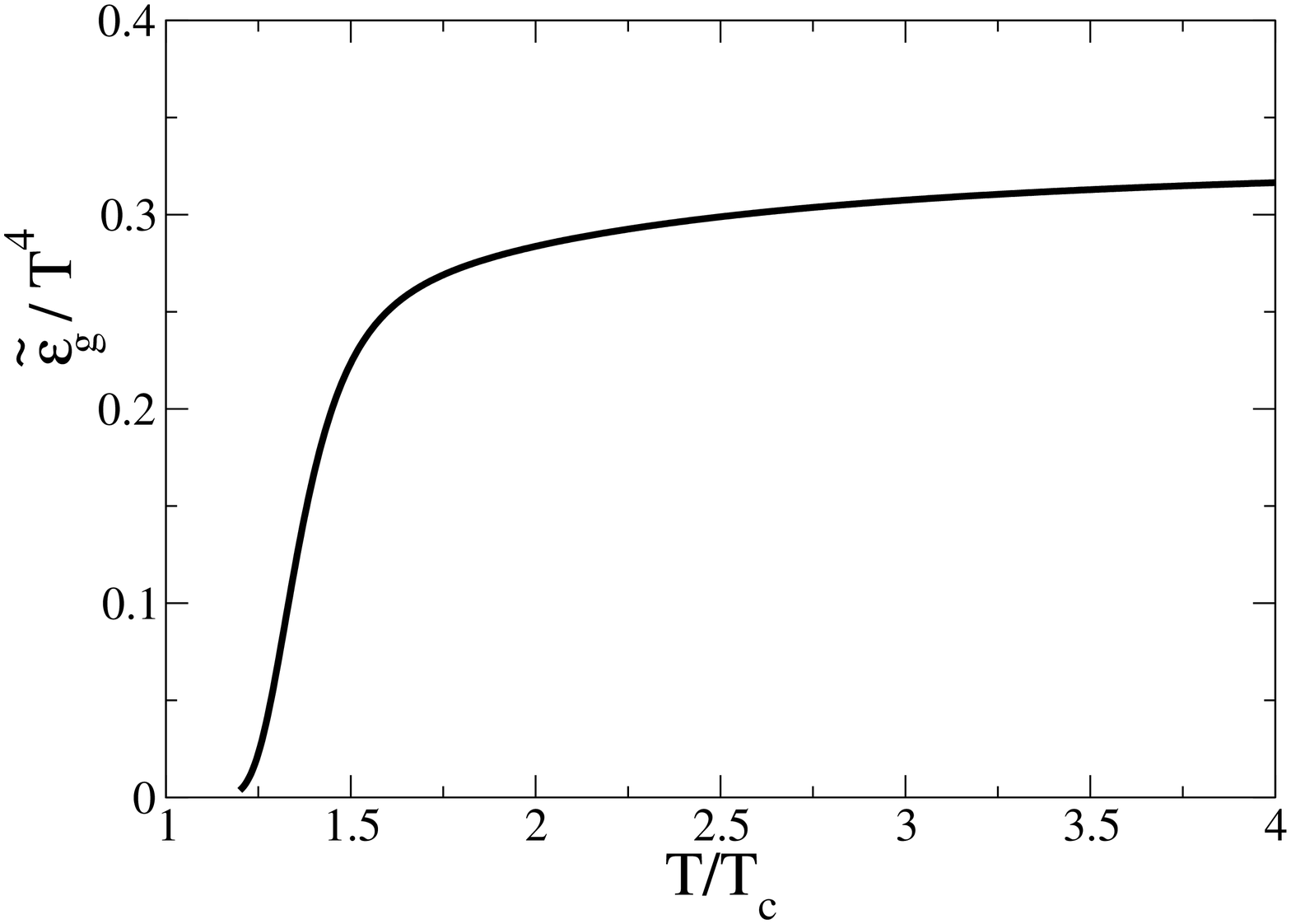}
\caption{Gluonic pressure (left panel) and energy density (right
panel) per degree of freedom as a function of the temperature.}
\label{gxdegree}
\end{figure}

In Fig.~\ref{gxdegree} we show the plot of the pressure  $\tilde
p_g$ and energy density $\tilde \epsilon_g$ per degree of freedom.
Note that  the energy density does not suddenly saturates at
$T\simeq1.2 T_c$ as observed in the lattice simulations. The reason
is that we have subtracted from the energy density the contribution
due to the gluon condensate.

A quantity that is relevant for the dynamical properties of the
system and gives a different measure of the deviation from the
conformal behavior is  the velocity of sound,   defined as, \be
c_s^2 = \frac{\partial p}{\partial \epsilon } \,.\label{vsoundg}\ee
For a conformal symmetric system the velocity of sound squared is
equal to $1/3$ and therefore deviation from this numerical value
indicate a breaking of conformal symmetry.

In our case we can define  the velocity of sound of the gluon
quasiparticles  by, \be c_{s,g}^2 = \frac{\partial p_g}{\partial
\epsilon_g}\,,\label{vsoundg2}\ee which is clearly different from
the total velocity of sound in Eq.~(\ref{vsoundg}). The two results
are shown in Fig.~\ref{gxdegree}; the dashed line (blue online)
corresponds to Eq.~(\ref{vsoundg}), whereas the full line (red
online)  corresponds to Eq.~(\ref{vsoundg2}).

\begin{figure}[!th]
\includegraphics[width=2.5in,angle=-0]{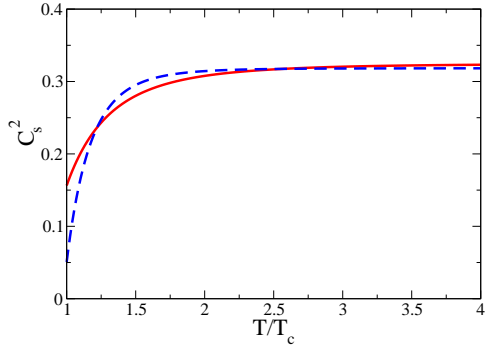}
\caption{(color online) The velocity of sound squared of the gluon
plasma as a function of the temperature. The full line (red online)
 corresponds to the velocity of sound for the gluon
quasiparticles. The dashed line (blue online)  corresponds to the
velocity of sound for the total system.} \label{soundglue}
\end{figure}

The two velocity of sound are quite similar except for $T \simeq
T_c$. Unfortunately this is the region where the errors in lattice
data are larger and one can assume that our result have at least a
20\%  error bar (for recent more accurate lattice data regarding the
velocity of sound in pure SU(3) see \cite{Gavai:2005da}).  Note that
the velocity of sound drives the hydrodynamical evolution of the
system produced in heavy-ion collisions and in particular of the
elliptic flow. Therefore a more detailed knowledge of the sound
velocity would be helpful in understanding the dynamical properties
of the system.

\section{Analysis of the Quark Gluon Plasma \label{QGP}}

Let us now  consider the case of the Quark-Gluon Plasma where gluons
as well as quarks are  dynamical degrees of freedom. As in the
previous Section we first consider the result of lattice simulations
and show that the gluon condensate is not sufficient to explain the
trace anomaly of the system. In analogy with Eq.(\ref{anomaly3}) let
us define $c_e^{m_q}(T)$ the ratio between the chromo-electric gluon
condensates at zero and finite temperature for finite quark mass
$m_q$. The pure gauge results,
 $c_e^{\rm qu}(T)$  corresponds to the limit $m_q \to \infty$.

It turns out that in full QCD with a quark mass $m_q= 0.1~T$,
$c_e^{m_q}(T)<c_e^{qu}(T)$ for any $T$
\cite{Novikov:1981xj,D'Elia:2002ck}. In the following we will refer
to $c_e^{m_q}(T)$ at $m_q= 0.1 ~T$ as $c_e^{\rm un}(T)$ and we will
assume that it is not strongly dependent on the actual value of the
quark mass.

\begin{figure}[!th]
\includegraphics[width=2.in,angle=-0]{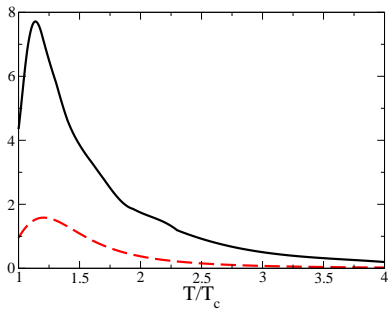}
\caption{(Color online) Interaction measure (black) and gluon
condensate (red) as a function of $T/T_c$.} \label{intmesure}
\end{figure}

Following the same approach of Sec.\ref{interaction}, in Fig.
\ref{intmesure} we compare the unquenched results for $(\epsilon
-3p)/T^4$ with the gluon condensate contribution, given by the
Equation (\ref{anomaly4}) with $<G^2>_0 \simeq 0.02$ GeV$^4$ and
$c_e^{\rm un}(T)$ obtained by interpolating the lattice QCD results
of Ref.~\cite{D'Elia:2002ck}.

In general, the trace anomaly has  a fermionic contribution
proportional to $<\bar q q>$. However, above the chiral symmetry
restoration temperature, that we consider coincident with  the
deconfinement temperature, one has $<\bar q q> = 0$.

\vspace{1cm}

In  full QCD we find a result similar to the one obtained in the
previous Section  for the gluon plasma: the gluon condensate is not
the only contribution to the trace anomaly. Therefore we rewrite
Eq.~(\ref{anomaly}) as follows:

\be <\Theta^{\mu}_{\mu}>_T = \epsilon - 3P\,= <G^2>_0-<G^2>_T  +
\Delta\Theta \label{anomaly5}\,, \ee where $\Delta\Theta$ represents
the contribution to the trace anomaly from gluonic, fermionic as
well as bosonic degrees of freedom.

\subsection{Quasiparticle model for the Quark-Gluon Plasma \label{unquenched}}

We shall assume that above the confining temperature the system
consists of fermionic quasiparticles with the quantum numbers of
$u$, $d$ and $s$ quarks, gluon quasiparticles and in-medium mesons
(or correlated pairs).

Concerning the fermionic sector,  we assume that the number of quark
(antiquark) degrees of freedom  $D_q$ ($D_{\bar q}$) is independent
of the temperature and  $D_q=D_{\bar q} = 18$.  This is motivated by
the fact that the baryon chemical potential is close to zero and
charm quarks are too massive to play a role in the temperature range
that we are studying.  The general form of the dispersion law for
in-medium fermions can be written as,
\begin{equation}
\omega_{\bar q}(k,T)\,=\,\omega_q(k,T)\,=\, \sqrt{k^2 + m(k,T)^2} +
\Sigma_R(k,T) \label{dispersion}\, ,\end{equation} and has been
studied in \cite{Mannarelli:2005pz,Kitazawa:2005mp}, where $m$ and
the self-energy $\Sigma_R$ have been  evaluated  taking into account
the interaction of the quasiparticles in the medium.

Here we will assume that for the relevant momenta, of the order of
the thermal momentum, we can neglect $m/k$ (see e.g.
\cite{Mannarelli:2005pz}) and treat the  term $\Sigma_R$ as a chiral
invariant  and temperature dependent effective mass $M(T)$. Quarks
are considered degenerate in mass because  $M(T)$  is generated by
the interaction of the quarks with the medium and for light quarks
this is independent of the value of the bare mass. From this
assumptions  the dispersion relation for $u, d$ and $s$
quasiparticles can be written as,
 \be \omega_{\bar
q}(k,T)\,=\,\omega_q(k,T)\,=\, k + M(T) \label{dispersion2}\, .\ee

The  structure  of the  in-medium correlated states as a function of
temperature is not easily to valuate. These string like states may
describe $\bar q q$ states as well as more exotic states
\cite{Shuryak:2004tx}. We will consider  that there are  $D_{b}(T)$
bosonic degrees of freedom and we expect that $D_{b}(T)$ increases
close to the $T_c$ and decreases with increasing $T$, because  for
asymptotic values of the temperature the system is made up of quarks
and gluons and   $D_{b}(T)$ must vanish.

In the following we will neglect, as a first approximation, the
effect on the thermodynamics quantities of the width of  the bosonic
states. Therefore we employ the dispersion law \be \omega_b(k,T) =
\sqrt{k^2 + M_b(T)^2}\,,\ee where $M_b(T)$ is the in medium mass of
the bosons. To evaluate  the dependence of the numerical results on
this parameter  we have changed the mesons masses in the range $M(T)
- 3 M(T)$ obtaining variation of less than $15 \%$ of our results.

Concerning  the gluonic sector we will employ the results obtained
in Section \ref{quenched}. Assuming that the effective number of
degrees of freedom and mass of gluons is given by the result
obtained in the quenched case  is clearly a rough approximation,
because it assumes that the effect of dynamical quarks on the gluon
quasiparticles is much smaller that the effect of the gluon medium.
Nonetheless we expect that in the unquenched case the effective
number of gluon degrees of freedom has a behavior similar to the one
obtained in the quenched case. Indeed we expect that $D_g(T)$
decreases with increasing temperature  going from 24 close to $T_c$
to $16$ at high temperatures, for the reasons explained in Section
\ref{mgdg}. Therefore we will  assume that $D_g(T)$ is given by the
same expression obtained in the quenched case.

Regarding the temperature dependence of the in-medium mass of the
gluons we expect in both the quenched and unquenched cases a similar
behavior. The reason is that at asymptotic high temperatures the
system has to become scale invariant and therefore the gluon mass
must decrease with increasing temperature. In the following we will
assume that the gluon mass in the unquenched case, $\hat M_g(T)$,
  is proportional to the  gluon mass in the quenched case,
$ M_g(T)$, that is shown in the left panel of Fig \ref{figglue}.
However, since the numerical value of the gluon mass in the
unquenched case can be different from the numerical value in the
quenched case, we write \be \hat M_g(T) = c M_g(T)\label{masshat}
\,,\ee where $c$ is a coefficient which parameterizes the variation
of the gluon mass in the medium when dynamical quarks are present.
In the numerical study we will consider the values $c=0.5,1,2$. The
corresponding dispersion law for gluon quasiparticles turns out to
be \be \hat\omega_g(k,T)=\sqrt{k^2 + \hat
M_g(T)^2}\label{gluondisp2}\,.\ee

Within these approximations,
 the expressions for the pressure, $p$,  and energy density, $\epsilon$, are given by
\begin{equation}
p=p_f + p_b + \hat p_g \label{Ptot}\end{equation}
\begin{equation}
\epsilon=\epsilon_f + \epsilon_b + \hat\epsilon_g + \epsilon_{\rm
con}^{\rm un}\, , \label{Etot}\end{equation} where the subscripts
$f, b$ and $g$ refer to fermionic, bosonic (in-medium mesons) and
gluonic degrees of freedom respectively and where \bea p_i(T) &=& T
 D_i\int \frac{d^3 k}{(2 \pi)^3} \log(1 \pm e^{-\omega_i/T})^{\pm 1}\,, \nonumber\\
\epsilon_i(T) &=&  D_i\int \!\!\!\frac{d^3 k}{(2 \pi)^3}
\frac{\omega_i}{e^{\omega_i/T}\pm 1} \label{pressenergy}\, ,\eea
with $i=f,b$; the sign $+$ ($-$) refers to fermions (bosons),
whereas the gluon contributions are given by  \bea \hat p_g(T) &=&
T D_g(T)\int \frac{d^3 k}{(2 \pi)^3} \log(1 - e^{-\hat\omega_g/T})^{-1} \nonumber\\
\hat \epsilon_g(T) &=&  D_g(T)\int \!\!\!\frac{d^3 k}{(2 \pi)^3}
\frac{\hat\omega_g}{e^{\hat\omega_g/T} - 1} \label{pressenergygg}\,
,\eea and the contribution of the gluon  condensate to the energy
density is given by, \be \epsilon_{\rm con}^{\rm un} = \frac{1}{2}
<G^2>_0 [1 - c_e^{\rm un}(T)] \,.\ee

In order to evaluate $M(T)$ and $D_b(T)$ we perform a simultaneous
fit of the lattice data of pressure and energy density of 3 flavors
quark matter  with  bare quark masses $m = 0.4\, T$  of
Refs.~\cite{Karsch:2003jg,Karsch:2000ps} as a function of the
temperature employing Eqs. (\ref{Ptot}) and (\ref{Etot}). For each
value of the temperature we consider the central value  of pressure
and energy density of the lattice data.  We will discuss the
dependence of our results on the numerical values of the lattice
data in the following. In \cite{Castorina:2005wi} we evaluated with
an analogous method $M(T)$ and $D_b(T)$ in the range $1.2-2. \,T_c$,
keeping the mass of the gluons and their effective number fixed;
here we extend such a study to a larger temperature range and
considering temperature dependent gluon mass and effective number of
degrees of freedom.

\begin{figure}[!th]
\includegraphics[width=3.in,angle=-0]{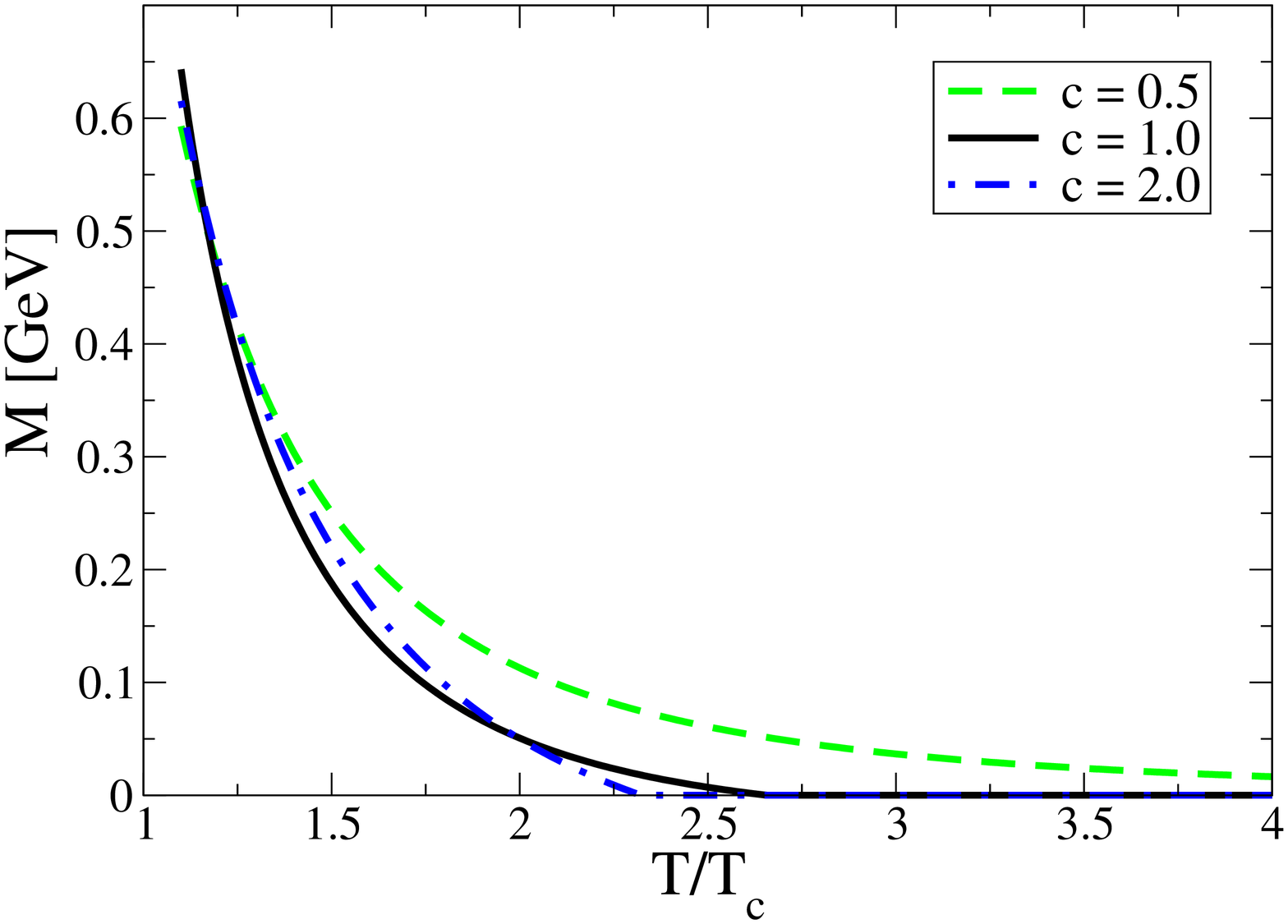}
\includegraphics[width=3.in,angle=-0]{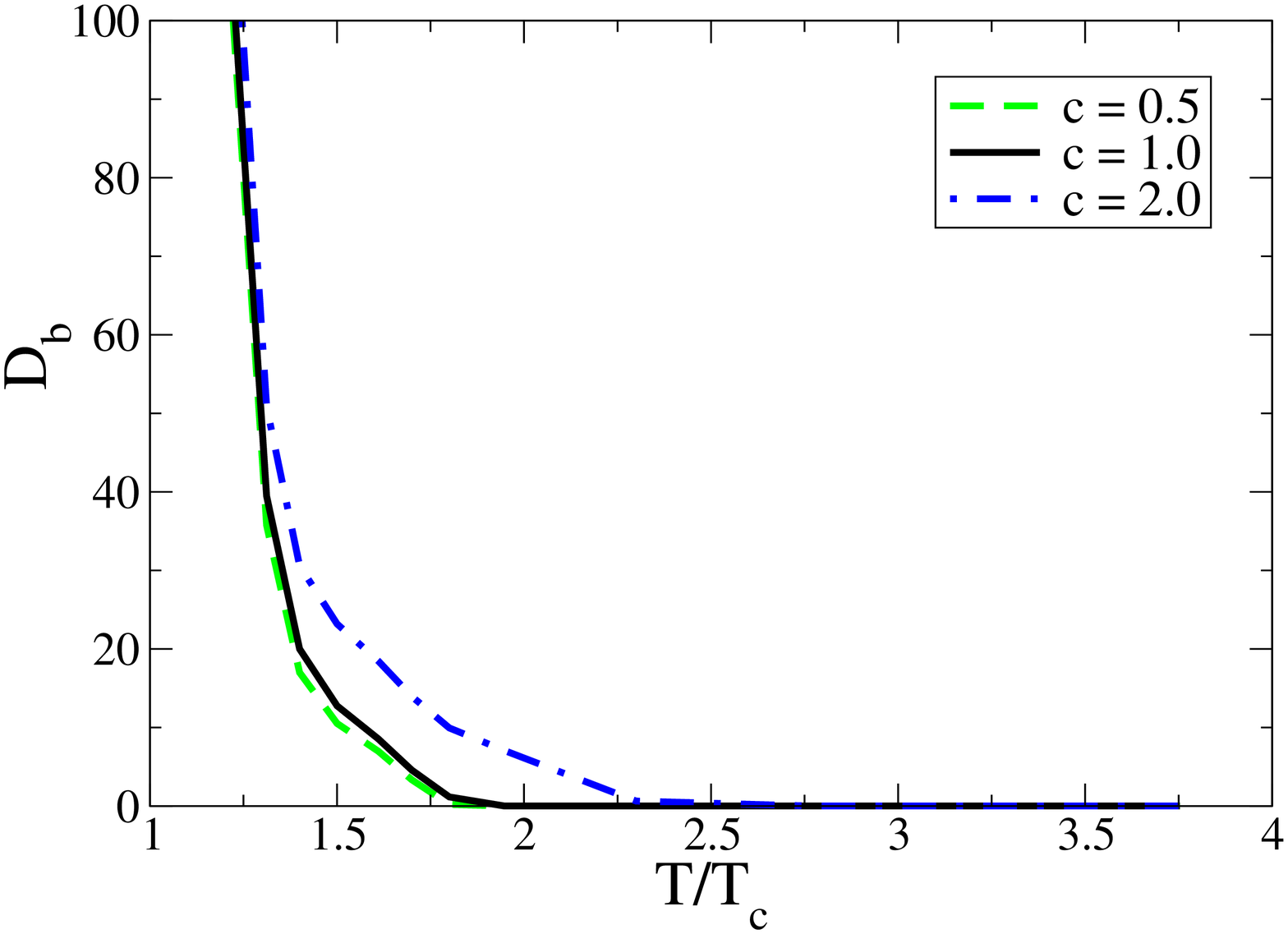}
\caption{(color online) Quasiparticle chiral mass (left panel) and
effective number of bosonic degrees of freedom (right panel) as a
function of the temperature for $T \simeq 1. - 4. T_c$. The gluon
effective mass employed in the evaluation of both the quantities is
given by Eq.~(\ref{masshat}) with the coefficient $c$ that takes the
values 0.5 for the dashed line (green online), 1 for the full line
(black online) and 2 for the dot-dashed line (blue online)
respectively.} \label{figNbM}
\end{figure}

In Fig.~\ref{figNbM} we show the  quasiparticle chiral mass (left
panel) and the effective number of bosonic degrees of freedom (right
panel)  as a function of the temperature for $T \simeq 1. - 4. T_c$.
 The various line correspond to different values of the parameter
$c$ in Eq.~(\ref{masshat}) that is the ratio between the gluon mass
in the quenched and unquenched cases. In the three cases considered
the chiral mass rapidly decreases with increasing temperature. This
behavior suggests that the breaking of scale invariance is mainly
due to the gluon mass and the gluon condensate.

Also the effective number of bosonic degrees of freedom  rapidly
decrease with increasing temperature. At temperature larger than
$1.5~T_c$, we find that less than $20$ bosonic modes are excited.
The divergence of $D_b(T)$ for temperature close to $T_c$ is due to
the same numerical problem that we have discussed in the previous
Section when we have presented  the results regarding the
temperature dependence of $D_g(T)$. However in this case the
exponential growth of $D_b$ begins at a value of the temperature
$\sim 1.2~T_c$. We have checked, that the numerical behavior below
$\sim 1.2~T_c$ is strongly dependent on the numerical value of the
gluon condensate and of  the energy density and pressure. On the
other hand for temperatures larger than $\sim 1.2~T_c$ our results
are basically almost independent  of the value of the lattice data.
Considering different values of energy density, pressure or gluon
condensate within the statistical error bars determines a variation
of less than 10\% of our results.

\begin{figure}[!th]
\includegraphics[width=2.5in,angle=-0]{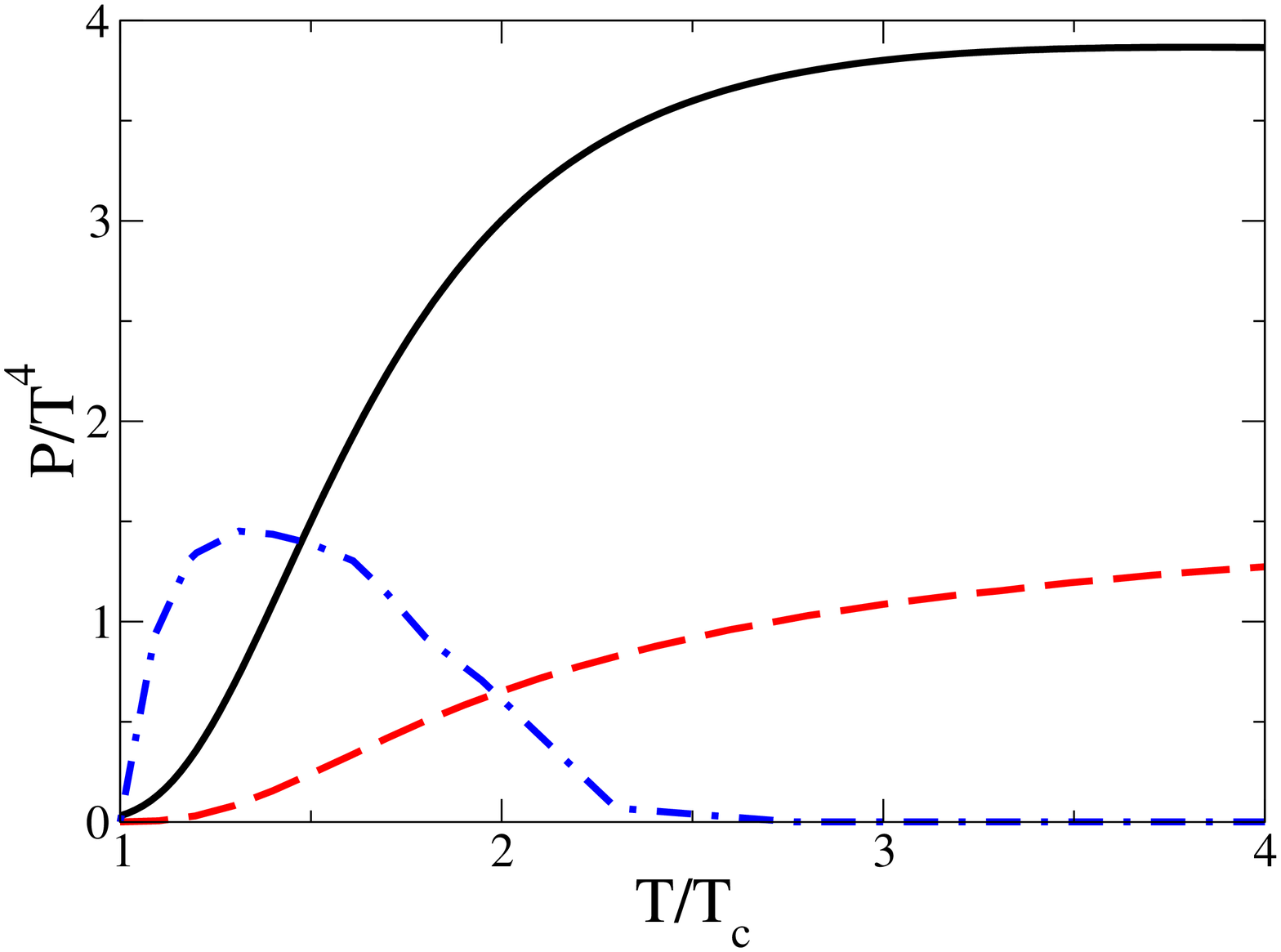} \hspace{1cm}
\includegraphics[width=2.5in,angle=-0]{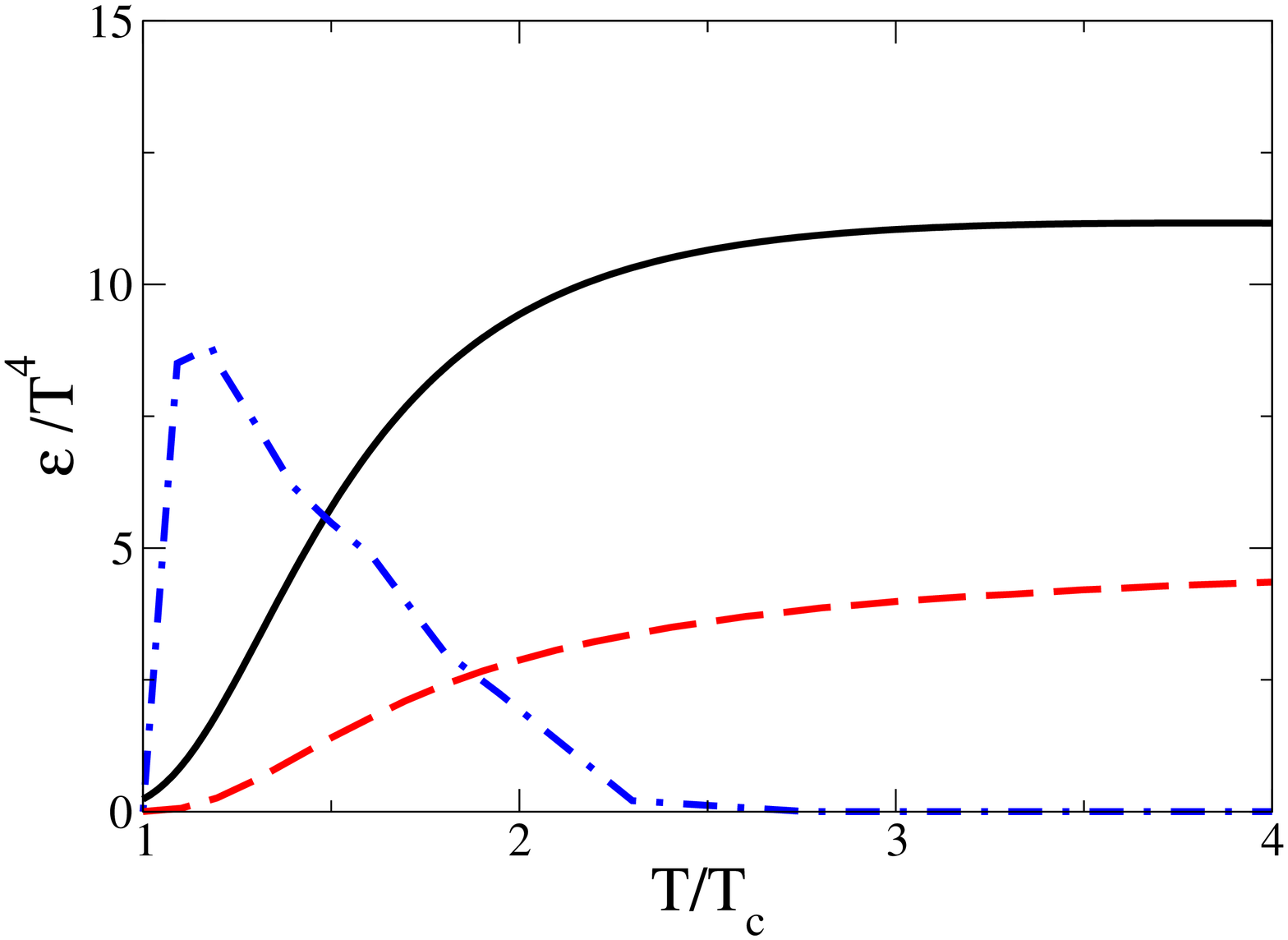}
\includegraphics[width=2.5in,angle=-0]{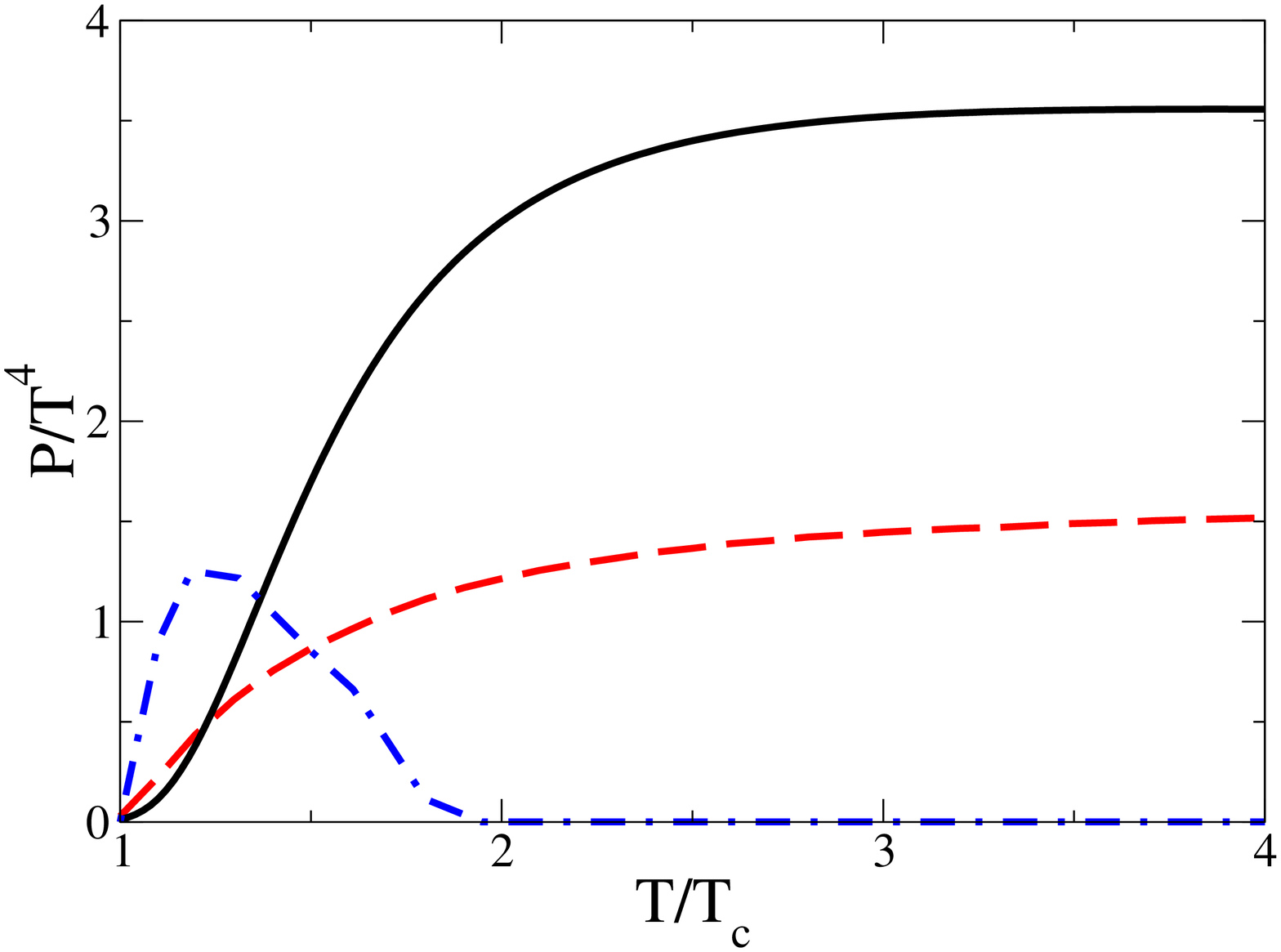} \hspace{1cm}
\includegraphics[width=2.5in,angle=-0]{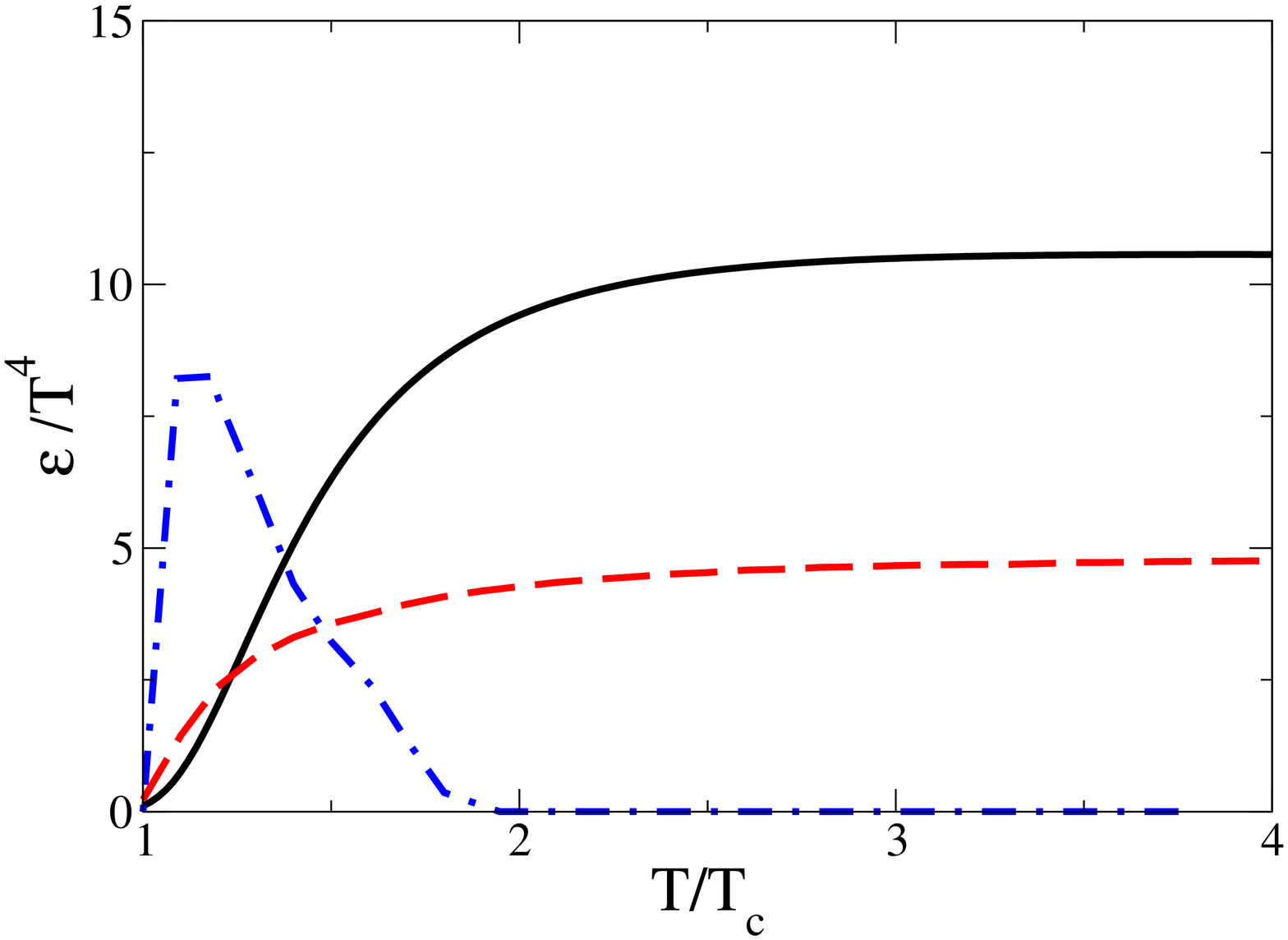}
\includegraphics[width=2.5in,angle=-0]{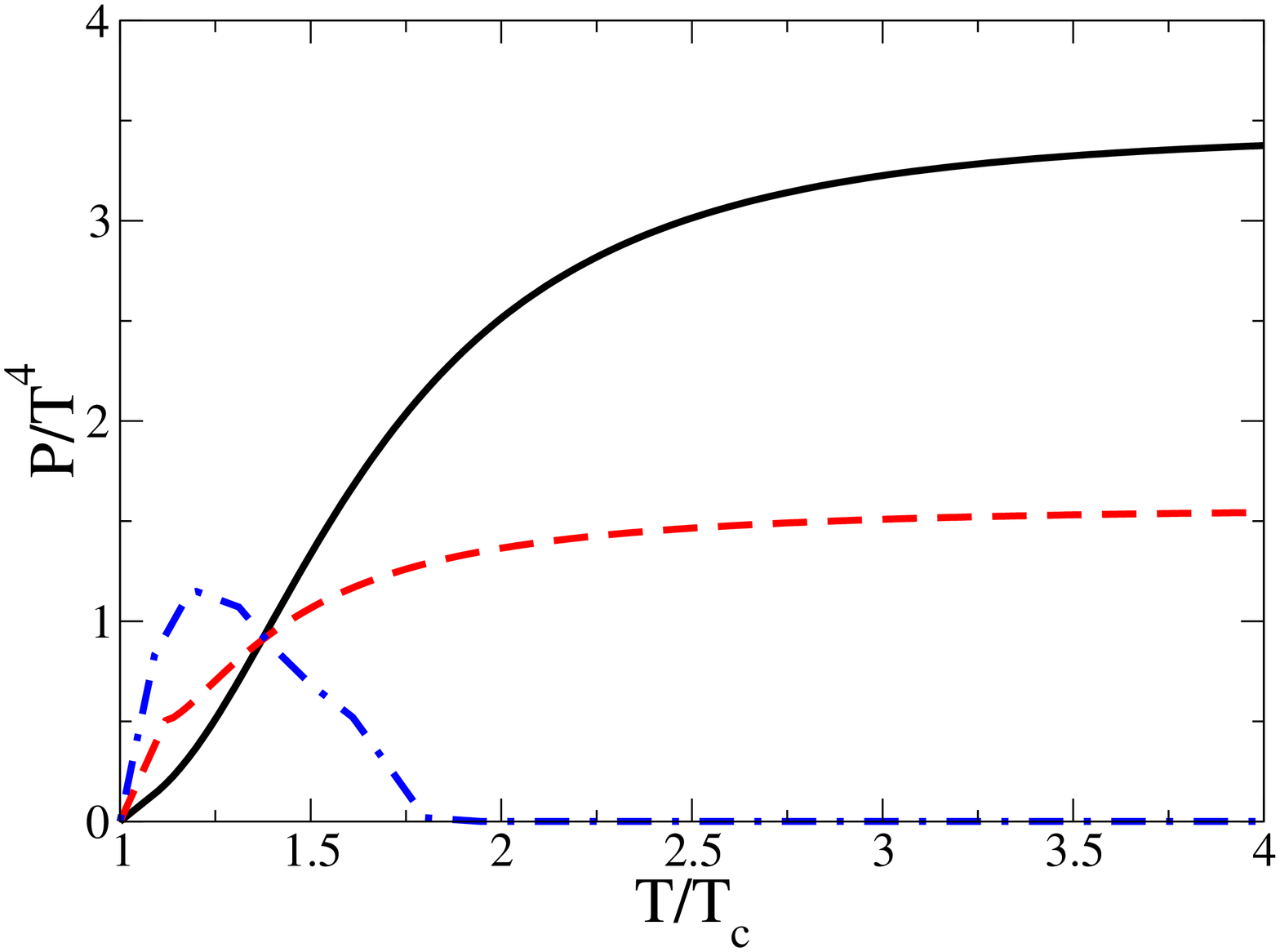} \hspace{1cm}
\includegraphics[width=2.5in,angle=-0]{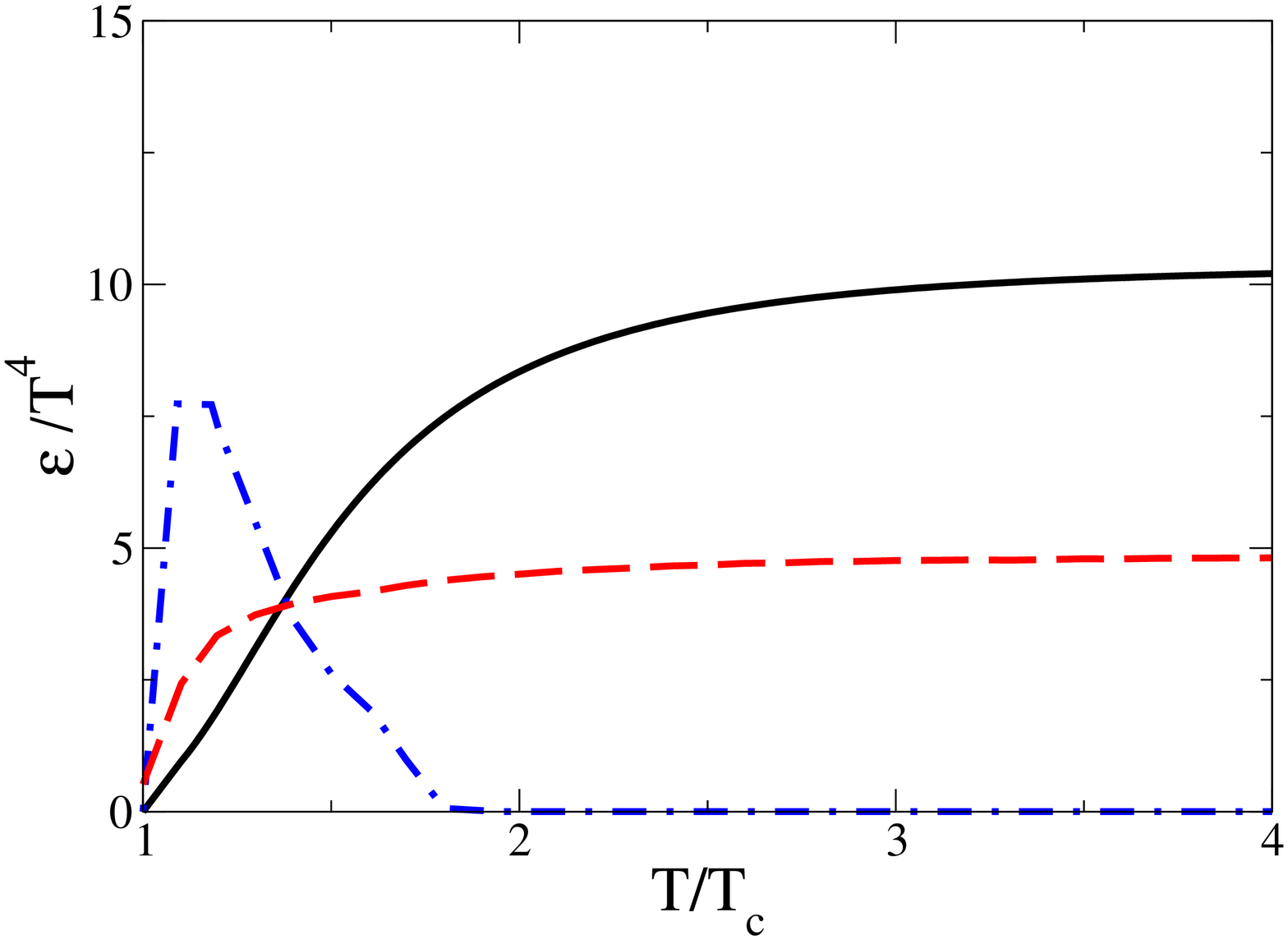}
\caption{ Contributions to the pressure (left panels) and energy
density (right panels) of fermionic (full black line), correlated
states (dot-dashed blue line) and gluonic (dashed red line) states.
The in-medium gluonic mass is given by Eq.(\ref{masshat}) where we
have employed different values of $c$.  From top to bottom we have
taken $c=0.5$, $c=1$ and $c= 2$ respectively.} \label{partialPE}
\end{figure}

In Fig.~\ref{partialPE} we show the contribution to the pressure and
energy density due to the
  fermions (full black line), gluons (dashed red line) and correlated  bosons (dot-dashed blue line)
quasiparticles employing a gluonic mass given by Eq.~(\ref{masshat})
with different values of $c$.

In all the considered cases  we find that at  high temperature the
dominant contribution to energy density and pressure is due to the
fermionic modes. However at sufficiently low temperature the
contributions of gluons and correlated pairs  become relevant. In
particular, the correlated states give a non negligible contribution
to energy density and pressure for temperature $\lesssim 1.5\, T_c$.
Unfortunately, as we have already stressed, for temperatures smaller
than $1.2\,T_c$ our results for the boson modes are not reliable.
However, it is reasonable that the qualitative behavior that we find, 
with a peak in energy density and in the pressure
at a certain value of the temperature, does not depend on our
approximations and is not an artifact of the numerical errors. The
reason is that close to $T_c$ energy density and pressure must be
zero, because of the normalization of the lattice data. On the other
hand  at large temperatures we expect that the system consists of
quarks and gluons,  the correlated pairs being melted and therefore
cannot contribute to the thermodynamical quantities. Then $p_b(T)$
and $\epsilon_b(T)$  must have a maximum for a certain value of the
temperature.

\begin{figure}[!th]
\includegraphics[width=1.95in,angle=-0]{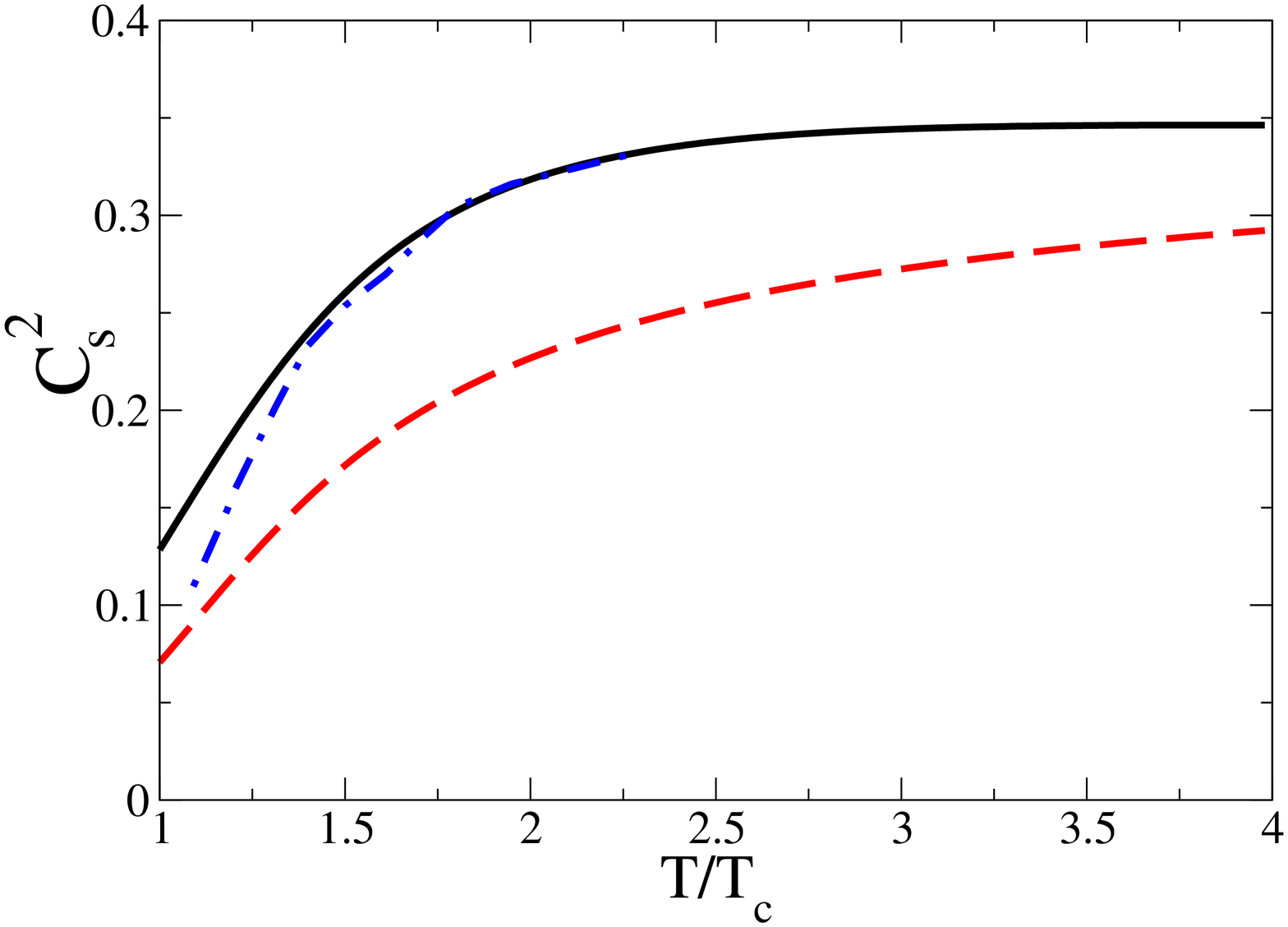} \hspace{0.5cm}
\includegraphics[width=1.95in,angle=-0]{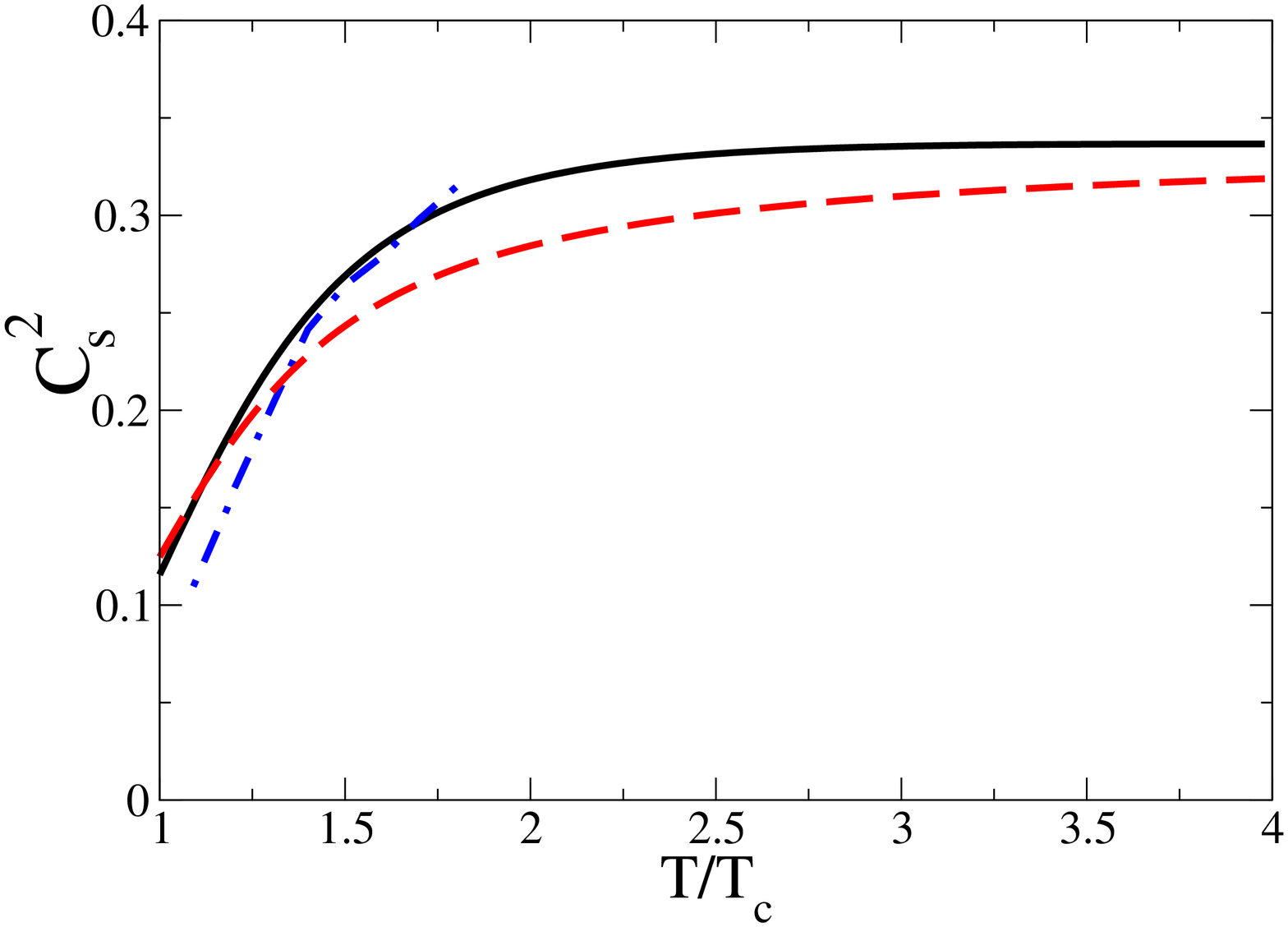}\hspace{0.5cm}
\includegraphics[width=1.95in,angle=-0]{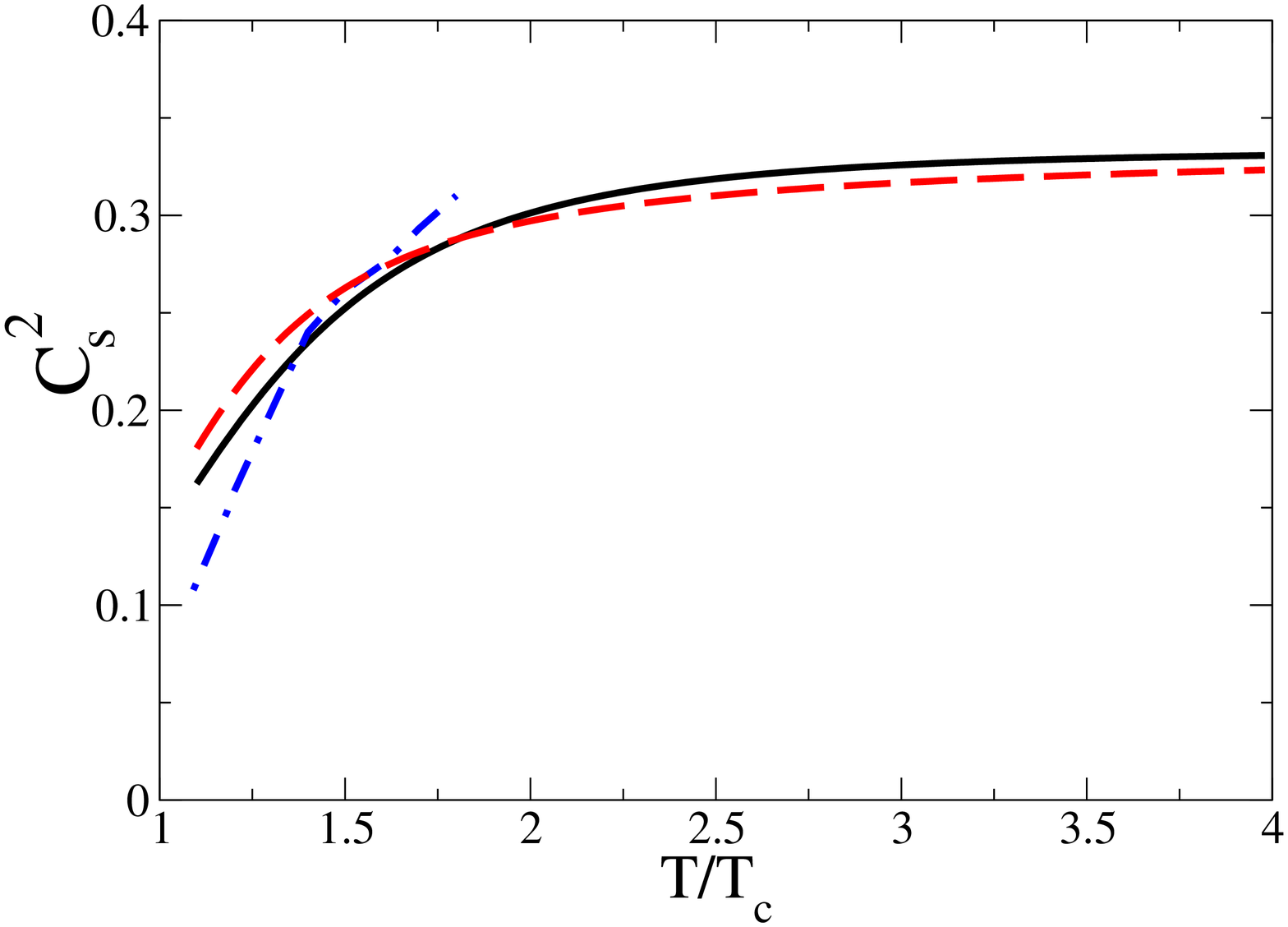}
\caption{(color online) Sound velocity of the various components.
Full line (black online) correspond to fermions, dashed line (red
online) corresponds to gluons and dot-dashed line (blue online)
corresponds to correlated pairs. From left  to right the three
panels correspond to the values $c=2,1,0.5$ respectively for the
in-medium gluonic mass given by Eq.~(\ref{masshat}).}
\label{partialcs}
\end{figure}

In Fig.~\ref{partialcs} we show the squared sound speed  of the
various quasiparticles for three different values of the parameter
$c$ defined in Eq.~(\ref{masshat}). At high temperatures the
fermionic quasiparticle have the largest velocity of sound, however
at moderate temperature it is not clear which component is dominant.
Note also that  fermionic and bosonic quasiparticles seem to be
``approximately conformal" for any value of $c$, in the sense that
the corresponding velocity of sound approaches the conformal values
already at temperatures of the order of $2 T_c$. In the case $c=2$
and $c=1$, gluons seem to be more scale dependent. However when we
reduce the gluon mass taking $c=0.5$ also gluons seem to approach
the conformal limit at a temperature close to $2 T_c$. Regarding the
correlated pairs, we have shown their behavior up to the temperature
where $D_b$ is different from zero.

\section{Conclusions and Outlooks \label{conclusion}}

In the present paper we have described the lattice QCD data of
pressure and energy density in terms of a quasiparticle model.  Such
description of lQCD data is not new. However our results show that
there are two dynamical ingredients, not taken into account in
previous analyses, which play an important role: the difference
between trace anomaly and gluon condensate and the survival of
correlated, string like, pairs for $T>T_c$, both suggested by
different lattice simulations.

The pure SU(3) gluon plasma lattice results  can be described in
terms of gluon condensate and gluonic quasiparticles whose mass
decreases with increasing  temperature and whose effective number of
degrees of freedom tends to  the expected value, 16, for large $T$.

Analogously,  numerical simulations results for QCD with finite
quark masses can be fitted  by the gluon condensate plus a mixture
of  gluonic and  fermionic quasiparticles and bosonic correlated
pairs.

Also in this case the effective number of bosonic degrees of freedom
and the in-medium masses decrease with increasing temperature. At
$T\simeq 1.5\,T_c$ only the correlated pairs corresponding to the
mesonic nonet survive and they completely disappear at $T \simeq 2\,
T_c$.

The temperature dependence of the sound velocity has been studied to
give indications for the hydrodynamical models of relativistic heavy
ion collisions.

The present analysis can give partial answers to important problems.
For example,  lattice data  show  a  deviations of the  pressure and
of the energy density of the system from the SB values of an ideal
gas of quarks and gluons even at temperatures  $T \ge 3\, T_c$. At
$T=3~T_c \simeq 500$ MeV the running coupling constant is  large and
the  deviations from the SB behavior can only be partially
understood with improved perturbative methods
\cite{Blaizot:2000fc,Blaizot:2003tw}. Our results show that at large
$T$ there are still effects due to quasiparticle masses that
determine variation from the SB values, whereas the contributions of
gluon condensate and of the correlated pairs turns out to be small.
The dominant contribution to the deviation from the ideal gas
results seems to be due to the mass of the gluons.

Our conclusions follow by considering a temperature independent
chromo-magnetic gluon condensate that has been evaluated in lattice
simulations only up to $T \simeq 2~T_c$. A gluon condensate growing
as $T^4$ for large $T$ suggested in Ref.~ \cite{Agasian:2003yw},
would modify the whole picture of the system for $T>> T_c$. However
it is not clear at which temperature the results of Ref.~
\cite{Agasian:2003yw} would be relevant. For this reason   we have
assumed a constant value of the chromo-magnetic gluon condensate,
extrapolating lattice data up to $4 T_c$.

Another limitation of the present study is the zero width
approximation of the correlated pair.  The effect of the width would
be to modify the expression of energy density and pressure of the
correlated pairs, but also to change in a self-consistent way  the
dispersion law of fermions
\cite{Schmidt:1990,Mannarelli:2005pa,Mannarelli:2005pz}. These
aspects and a more refined treatment of the gluon condensate will be
analyzed in a forthcoming paper.

{\bf Acknowledgements}

The authors thank D.~Cabrera, G.~Nardulli, R.~Rapp,  H.~Satz and H.~van Hees  for
useful discussions. The work of MM has been supported by the ``Bruno
Rossi" fellowship program. MM  acknowledges the hospitality of the
Perimeter Institute during the completion of this work. This
research was supported in part by the U.S.~Department of Energy
under cooperative research agreement \#DF-FC02-94ER40818.

\appendix
\section{Thermodynamics consistency}

In order to describe the pressure and energy density of the Gluon
Plasma in Section \ref{quenched} and of the Quark-Gluon Plasma  in
Section \ref{unquenched} we have employed a quasiparticle expression
for the various thermodynamical quantities. Since we have assumed
that  some parameters are temperature dependent the thermodynamics
consistency of our results must be carefully checked. In general we
can write the relation between pressure and energy density as
follows, \be T \frac{d p}{dT} = p + \epsilon + C_r\, , \ee where the
correction $C_r$ depends on the temperature. The thermodynamics
consistency requires that \be
 C_r=0 \,.
\ee

The correction  to the thermodynamical relation for the Gluon Plasma
is due to the temperature dependence of the parameters $M_g(T)$ and
$D_g(T)$ and to the introduction of the gluon condensate and is
given by, \be C_r^{\rm qu}(T)= T M_g \frac{d M_g}{d
T}\frac{D_g}{2\pi^2} \int dk k^2 \frac{1}{1- \exp{(\omega_g/T)}}
\frac{1}{\omega_g} + \frac{T}{D_g}\frac{dD_g}{dT} p_g- \epsilon_{\rm
con}^{\rm qu} \,.\label{crg}\ee Requiring that $C_r^{\rm qu}(T)=0$,
is equivalent to requiring that $D_g$ satisfies the differential
equation  \be A \frac{dD_g}{dT}+ B D_g + C = 0 \, , \label{diffg}\ee
where the coefficients can be obtained from Eq.~(\ref{crg}) and read
\bea A &=& T \tilde p _g \\ B &=& T M_g \frac{d M_g}{d T}{2\pi^2}
\int dk k^2 \frac{1}{1- \exp{(\omega_g/T)}}
\frac{1}{\omega_g} \\
C &=& -\epsilon_{\rm con}^{\rm qu} \,.\eea Substituting the values
of $D_g(T)$ and of $M_g(T)$, obtained fitting the lattice data of
energy and pressure,   in the previous equations, represents a
self-consistent check of our result.  We find that the differential
equation (\ref{diffg}) is satisfied with a good accuracy and the
corresponding correction $C_r^{\rm qu}$ is of the same order of the
error in the lattice data. Therefore in our method the differential
equation (\ref{diffg}) turns out to be selfconsistently satisfied
within the range of values of the statistical errors in lattice
data.

We can treat the correction to the thermodynamical equation   for
the Quark-Gluon Plasma in a similar way. In this case such a
correction is due to the temperature dependence of the parameters
$M(T)$, $D_b(T)$, $M_g(T)$, $D_g(T)$ and to the introduction of the
gluon condensate.  In this case we obtain, \bea C_r^{\rm un}(T) &=&
- T \frac{dM}{dT}\frac{2 D_q}{2\pi^2} \int dk k^2 \frac{1}{1 +
\exp{(\omega_q/T)}}  \nonumber
\\
& &+ T 4M\frac{dM}{dT}\frac{D_b}{2\pi^2} \int dk k^2 \frac{1}{1 -
\exp{(\omega_b/T)}} \frac{1}{\omega_b} +
\frac{T}{D_b}\frac{dD_b}{dT} p_b\nonumber\\& & + T \hat M_g \frac{d
\hat M_g}{d T}\frac{D_g}{2\pi^2} \int dk k^2 \frac{1}{1-
\exp{(\hat\omega_g/T)}} \frac{1}{\hat\omega_g} +
\frac{T}{D_g}\frac{dD_g}{dT} \hat p_g- \epsilon_{\rm con}^{\rm un}
\,.\eea  Also in this case the thermodynamical consistency can be
cast as a  differential equation for the effective number of degrees
of freedom. Assuming that the dependency of $D_g$ on the temperature
is the same as in the quenched case we obtain a differential
equation for $D_b$ of the form \be A \frac{dD_b}{dT}+ B D_b + C = 0
\, , \label{diff}\ee where  the coefficients $A,B$ and $C$ are given
by \bea A &=& -
\frac{T}{2\pi^2} \int dk k^2 \log[1 - \exp{(-\omega_b/T)}]\, \\
B &=& - \frac{T}{2\pi^2} \int dk k^2 \frac{1}{-1 +
\exp{(\omega_b/T)}} \frac{4M}{\omega_b}\frac{dM}{dT} \, \\ C &=&
-T\frac{2 D_q}{2\pi^2} \int dk k^2 \frac{1}{1 + \exp{(\omega_q/T)}}
\frac{dM}{dT}\\ && + T \hat M_g \frac{d \hat M_g}{d
T}\frac{D_g}{2\pi^2} \int dk k^2 \frac{1}{1- \exp{(\hat\omega_g/T)}}
\frac{1}{\hat\omega_g} + \frac{T}{D_g}\frac{dD_g}{dT} \hat p_g-
\epsilon_{\rm con}^{\rm un} \,. \eea

Substituting the values of $D_g(T)$, $M_g(T)$, $D_b(T)$ and $M(T)$
obtained fitting the lattice data of energy and pressure,  we find
that the differential equation (\ref{diff}) is satisfied with a good
accuracy. Also in this case the correction $C_r^{\rm un}$ is of the
same order of the error in the lattice data.

\end{document}